\documentclass[aps,twocolumn,prd,superscriptaddress]{revtex4-2}

\usepackage[utf8]{inputenc}

\usepackage{mathtools}
\usepackage{amsfonts}
\usepackage{mathrsfs}
\usepackage{bbm}
\usepackage{slashed}
\usepackage{tensor}
\usepackage{listings}
\usepackage{graphicx}
\usepackage{color}
\usepackage[dvipsnames]{xcolor}
\usepackage{array}
\usepackage[abs]{overpic}

\usepackage{placeins}

\usepackage{makecell}
\usepackage{subcaption}

\usepackage{xspace}
\usepackage{siunitx}
\usepackage{xfrac}
\usepackage{hyperref}
\usepackage[nameinlink]{cleveref}
\usepackage{appendix}
\usepackage{units}

\usepackage{xifthen}
\usepackage{xcolor}
\hypersetup{
	colorlinks,
	linkcolor={red!75!black},
	citecolor={blue!75!black},
	urlcolor={blue!75!black}
}

\usepackage{booktabs}
\usepackage{multirow}

\newcolumntype{C}{>{$}c<{$}}
\AtBeginDocument{
	\heavyrulewidth=.08em
	\lightrulewidth=.05em
	\cmidrulewidth=.03em
	\belowrulesep=.65ex
	\belowbottomsep=0pt
	\aboverulesep=.4ex
	\abovetopsep=0pt
	\cmidrulesep=\doublerulesep
	\cmidrulekern=.5em
	\defaultaddspace=.5em
}

\usepackage[edges]{forest}
\usetikzlibrary{arrows.meta}

\newcolumntype{L}{>{\centering\arraybackslash}m{3cm}}

\graphicspath{{./figures/}}

\newcommand{\gettitle}{Local Discontinuous Galerkin for the Functional Renormalisation Group}

\newcommand{\getHeidelbergAffiliation}{\affiliation{Institut f{\"u}r Theoretische Physik, Universit{\"a}t Heidelberg, Philosophenweg 16, 69120 Heidelberg, Germany}}
\newcommand{\getDarmstadtAffiliation}{\affiliation{Institut f\"ur Kernphysik, Technische Universit\"at Darmstadt, D-64289 Darmstadt, Germany}}
\newcommand{\getEMMIAffiliation}{\affiliation{ExtreMe Matter Institute EMMI, GSI, Planckstr. 1, 64291 Darmstadt, Germany}}

\hypersetup{
	pdftitle={\gettitle},
	pdfauthor={Ihssen, Pawlowski, Sattler, Wink},
	pdfkeywords=
	{functional renormalisation group} {effective potential} {LDG} {DUNE} {Discontinuous Galerkin},
	bookmarksopen=true,
	bookmarksopenlevel=2,
	bookmarksnumbered=true
}

\begin{document}

\title{\gettitle}

\author{Friederike Ihssen}\getHeidelbergAffiliation
\author{Jan M. Pawlowski}\getHeidelbergAffiliation\getEMMIAffiliation
\author{Franz R. Sattler}\getHeidelbergAffiliation
\author{Nicolas Wink}\getDarmstadtAffiliation

\begin{abstract}
We apply the Local Discontinuous Galerkin discretisation to flow equations of the O(N)-model in the Local Potential Approximation. The improved stability is directly observed by solving the flow equation for various $N$ and space-time dimensions $d$. A particular focus of this work is the numerical discretisation and its implementation. The code is publicly available, and is explained in detail here. It is realised as a module within the high performance PDE framework DUNE.
\end{abstract}

\maketitle

\section{Introduction}

Our modern understanding of equilibrium Quantum Field Theories (QFT) is in part based on the use of the Renormalisation Group (RG). Understanding and monitoring the successive emergence of the underlying physics and mechanisms in a given theory along a coarse graining scale trajectory is key to the resolution of the latter. A prominent example is the universal scaling physics in the vicinity of second order phase-transitions. However, the RG, and in particular is modern formulation in terms of the Functional Renormalisation group (fRG) is by no means restricted to the physics of phase transitions, but by now is a fully developed non-perturbative tool for resolving all aspects of quantum field theories. 

For example, in the context of the ongoing quest for the phase structure of QCD, the RG excels at connecting low energy effective theories to the fundamental degrees of freedom in QCD, see e.g.~\cite{Dupuis:2020fhh, Fu:2019hdw, Fu:2022gou}. In particular the fRG for the one-particle irreducible effective action as introduced in~\cite{Wetterich:1992yh} has found widespread use in many applications ranging from quantum gravity at the smallest momentum scales over Standard model and Beyond Standard Model physics and QCD at high energy scales, to statistical physics and condensed matter systems. The uniform applicability of the fRG in such a wide range of theories allows for a natural evaluation of their interrelations and similarities. This goes hand in hand with the identification of universal challenges. 

A prominent universal challenge concerns the task of charting phase structure of theories with competing order effects, (multi-)critical endpoints and lines of first and second order as well as crossover regimes. In short, an even qualitative access to some of the most eminent physics challenges in this context require a firm quantitative grip on the non-trivial dynamics in the respective regimes. For the challenge of charting the location of phase boundaries as well as their nature, the task turns out to be two-fold, with both aspects being deeply interconnected on a technical level:

The first task concerns first order phase transitions, which have recently been linked to shocks forming in field space \cite{Grossi:2019urj, Aoki:2017rjl}. This entails, that the reliable prediction of first order phase transitions or their absence can only be reliably predicted within a numerical implementation of the fRG approach that captures the emergence and evolution of shocks.

The second task concerns competing order effects, where even small changes in the flow can have qualitative impact on the resulting ground state. Accordingly, this challenges require both the resolution of momentum dependences of higher scattering  kernels and an quantitative resolution of full field dependencies in the fRG. 

Consequently, for an access to the phase structure of generic theories ranging from highest to lowest momentum momentum scales within a numerical  fRG approach, a numerically fully reliable and versatile framework is required, that captures the intricacies discussed above. In this paper we significantly add to this task, initiated in \cite{Grossi:2019urj} and built upon in \cite{Grossi:2021ksl, Koenigstein:2021syz, Koenigstein:2021rxj, Steil:2021cbu, Stoll:2021ori, Ihssen:2022xjv}, by introducing Local Discontinuous Galerkin (LDG) methods to the fRG approach at the example of fRG flows for a real scalar field. The respective code is available as an open access repository and can also be implemented for generic theories after respective modifications.  

Partial Differential equations (PDE) arise naturally in the fRG in the context of phase transitions. A common truncation scheme in this case is the derivative expansion, for a recent review see~\cite{Dupuis:2020fhh} and reference therein. In this truncation scheme the effective action is expanded in gradients of the field. To leading order, the dynamics is then encoded by the effective potential, whose flow is governed by a second order PDE.

In \cite{Grossi:2021ksl} we have introduced the Direct Discontinuous Galerkin (DDG), to handle second order derivatives in the resulting fRG flow equations. However, this method introduces large diffusive contributions. This problem is particularly severe in the vicinity of first order phase transitions. Due to the peculiar nature of the underlying equations, we suspect that, as a result, the method lacks proper convergence due big diffusive contributions. These shortcomings can be overcome within Local Discontinuous Galerkin (LDG) methods~\cite{Cockburn98thelocal}. They properly handle the more subtle structure of the second order derivatives in these equations, leading to enhanced numerical stability. As the implementation of these schemes becomes increasingly complex, a major part of the paper focuses thereon. In particular, we introduce an open source version of the code, published at \href{https://github.com/satfra/dune-FRGDG}{https://github.com/satfra/dune-FRGDG}, combined with a detailed explanation of the code in this paper. This enables interested readers to use these methods themselves without having to start from scratch. The code is written as a module for the high performance PDE framework DUNE~\cite{dune24:16}.

We start by recapitulating the fRG in \Cref{sec:FRG}. This is followed by an introduction of the LDG scheme for fRG equations in \Cref{sec:Numerics}. The LDG formulation is then applied to a zero dimensional QFT in \Cref{sec:HarmO} and to an O(N)-model in $d=0\ldots 4$ dimensions in \Cref{sec:ON}. Finally, \Cref{sec:numFrame} showcases the implementation of the LDG code in these models, and the code are publicly accessible on GitHub~\cite{LDGGit}.

\section{Functional Renormalisation Group}\label{sec:FRG}

In this Section we only briefly summarise the fRG for the readers convenience, more detailed accounts and references can, for example, be found in~\cite{Berges:2000ew, Pawlowski:2005xe, Gies:2006wv, Braun:2022mgx}. For simplicity we only consider a single component real scalar field $\varphi\in\mathbbm{R}$ with the generating functional $Z[J]$, 
\begin{align}\label{eq:Z-J}
	Z[J] = \int d\varphi \, e^{-S[\varphi] + J \varphi }\,,
\end{align}
with the Euclidean classical action 
\begin{align}\label{eq:Scl}
	S[\varphi] =\int d^d x\,\left[ (\partial_\mu \varphi)^2 +V_\textrm{cl}(\varphi^2/2)\right]\,,
\end{align}
with a quadratic classical dispersion $-\partial_\mu^2$ and the classical $\phi^4$ potential $V_\textrm{cl}$. An infrared regularised version is obtained from \labelcref{eq:Z-J} by adding a momentum dependent mass term to the dispersion of the scalar field, $-\partial_\mu^2\to -\partial_\mu^2 +R_k(-\partial_\mu^2)$. While $R_k(0)\propto k^2 $ is an infrared mass that suppresses quantum fluctuations, ultraviolet quantum fluctuations are integrated out, $R_k(p^2/k^2\to \infty)=0$. The modification of the dispersion is included with 
\begin{align}
	Z_k[J] = e^{-\Delta S_k\left[{\scriptstyle\frac{\delta}{\delta J}}\right]}Z[J]\, ,
	\label{eq:modGenFunc}
\end{align}
with the regulator term
\begin{align}
	\Delta S_k[\varphi] =  \frac{1}{2} \int \mathrm{d}^d x\, \varphi R_{k} \varphi\, .
\end{align}
The regulator properties discussed below  \labelcref{eq:Scl} ensure that $Z_{k\to 0}=Z$, the full generating function in \labelcref{eq:Z-J}, and tends toward a free (Gau\ss ian) theory for $k\to\infty$. 

In this work, we chose a flat (Litim) regulator, which allows us to perform the momentum integral analytically, 
\begin{align}
\label{eq:litim_reg}
	R_k(p) = (k^2 - p^2) \Theta(k^2-p^2)\, ,
\end{align}
and regulates only infrared momentum fluctuations below $k^2$. While the generating functional $Z_k[J]$ generates the full correlation functions including their disconnected parts, its logarithm generates connected correlation functions,  
\begin{align}
	W_k[J] = \ln Z_k[J]\,,
\end{align}
the scale dependent Schwinger functional. A further reduction of redundancies is achieved by considering the the Legendre transform of $W_k[J]$, the one-particle irreducible effective action, 
\begin{align}
	\label{eq:modLegendreFRG}
	\Gamma_k[\phi] = \sup\limits_{J} \bigg(W_k[J] - \int \mathrm{d}^d x\, J \phi \bigg) - \Delta S_k[\phi]\, .
\end{align}
Infinitesimal changes of the theory in terms of the effective action are comprised by the logarithmic scale derivative of \labelcref{eq:modLegendreFRG},  the Wetterich equation~\cite{Wetterich:1992yh}, 
\begin{align}
\label{eq:wetterich}
	\partial_t \Gamma_k[\phi] 
	= \frac{1}{2} \mathrm{Tr}\, \frac{1}{\Gamma_k^{(2)} + R_{k} } \partial_t R_{k}
\, ,
\end{align}
with the RG-time $t = -\ln\frac{k}{\Lambda}$ and the trace simply is a momentum integration for the present example of a real scalar field. Please note the additional minus sign in the definition of the RG-time in comparison to the standard notation. This choice is made to ensure a positive time evolution, which is natural in most computational approaches.

\section{Numerics}\label{sec:Numerics}

Originally developed for purely convective hyperbolic equations, the DG method has later been adapted for PDEs which are convection-dominated but have a diffusive component. There are several approaches to do that:

In \cite{Grossi:2021ksl} a direct DG extension has been applied to the problem of a quark-meson model, which gets a non-linear diffusive component through the massive mode that arises when the theory exhibits spontaneous symmetry breaking. The direct extension consisted of obtaining necessary derivative terms directly from the solution. This means that for a numerical approximation on a grid of $K$ cells and an expansion in a polynomial basis of order $P$ within each cell, to wit,
\begin{align}\label{eq:polybasis}
	{\bf w}_h^k (t,x) = \sum_{p=1}^{P+1} {\bf w}_p^k(t) \psi_p(x)\,,
\end{align}
the derivative $\partial_x{\bf w} (t,x)$ is given by
\begin{align}
	\partial_x{\bf w}_h^k (t,x) = \sum_{p=1}^{P+1} {\bf w}_p^k(t) \partial_x\psi_p(x)\,.
\end{align}
However, this direct definition of a numerical derivative is problematic. In \cite{zhang11} it has been shown that this formulation yields errors of O(1) even if it is stable. In the case of \cite{Grossi:2021ksl}, this method has also shown itself to be unstable in the vicinity of the first order regime of the quark-meson model.

To remedy the shortcomings of this naive extension, we will use the local discontinuous Galerkin method \cite{Cockburn98thelocal}. It has become one of the most widely used extension of the DG method in this direction.
In the remainder of this Section we will review the local discontinuous Galerkin method from \cite{Cockburn98thelocal}, as well as adapt it, using some of inherent properties of the fRG equations and the physics contained within. We remark that the LDG method is particularly suited to solve convection-diffusion equations with possibly discontinuous solutions, retaining its stability and convergence \cite{GeneralShu}.

\subsection{Assumptions on the form of the equations}

As a guiding example, let us examine the flow equation for an O(N) system, which we will keep in mind when explaining the LDG method.
An inherent property of the flow is its dependence on the second derivative of the effective action. This translates to a system of equations containing convection, as well as diffusion.
Instead of directly looking at the effective potential $V(x)$ of the theory, we take its first derivative and note that the two unknowns of the system are linked by another derivative
\begin{align}
	u = \partial_x V \,,\quad \mathrm{and} \quad v =  \partial_x^2 V \,.
\end{align}
Thus we introduce $\partial_x u$ as a further unknown, apart from $u$
\begin{align}\label{eq:dxu}
\partial_x u = v  \,.
\end{align}
Having established these definitions we allow for the following general form of the equations
\begin{align}\label{eq:originalEq} \notag
\partial_t u - \partial_x F_t(u,v) &=0 \,,\\[1ex]
\partial_t v - \partial_x \big( G_t(u,v) +  a_t(s) \partial_x s\vert_{s=u + b(x) v} \big)&= 0 \,,
\end{align}
where $u = u(t,x)$, $v=v(x,t)$ are the sought after solutions and we have some initial conditions at $t=0$ which we evolve in time.
This specific form is motivated by the expressions derived in \Cref{sec:HarmO} and \Cref{sec:ON}. It is given at this point to construct the numerical framework in a general way.
$u(t,x)$ is the solution to a purely convective equation with a Lipschitz conservative, RG-time-dependent flux $F_t$, whereas the equation for $v(t,x)$ has a convective (Lipschitz) flux $G_t$ and allows for a diffusive term linear in $\partial_x v$, which inherently depends on the combination $s=u+ b(x) v$, with a positive prefactor $a_t$. For simplicity we drop the $(\cdot)_t$ subscript from now on, which indicates the RG-time dependence of the terms.
The diffusive term in $\partial_x v$ prevents the use of direct DG methods discussed previously in \cite{Grossi:2019urj,Grossi:2021ksl}, since there is no sensible formulation of a numerical flux for higher derivative terms. In general the direct DG method is still applicable but has possibly non-negligible errors, as discussed above, if its contribution to the numerical flux is small.

The LDG method, as developed in \cite{Cockburn98thelocal}, presents a formulation for this additional diffusive flux, by solving a second, stationary equation, which will be discussed in \Cref{sec:numStat}. Afterwards, \Cref{sec:numflux} will discuss the introduction of the appropriate numerical fluxes.
Note, that introducing $v = \partial_x u$ as an independent unknown does not remedy the problem of having a second-order PDE, but it enables us to fit the framework given by the LDG method and thus construct a convergent and stable scheme for the fRG equations of such systems. The price to pay is a potentially reduced order of convergence due to promoting a dependent quantity into an independent one. However, this effect is practically mitigated by the high order of accuracy the LDG method. Its error is of order $O(\Delta x)^{P+\frac{1}{2}}$, where $\Delta x$ is the grid spacing and $P$ is the maximal order of the trial functions.

\subsection{Deriving the stationary Equation}\label{sec:numStat}

We will now present the LDG formulation for the system given in \labelcref{eq:originalEq}. The main idea here is to introduce a stationary equation which captures the additional derivative term by rewriting it as $q = \sqrt{a(u + b(x)v)}  \  \partial_x( u +  b(x) v)$
\begin{align}\label{eq:DGForm}
\partial_t u + \partial_x ( F(u, v)) &= 0 \,,\notag \\[1ex]
\partial_t v + \partial_x \big(G( u, v) -\sqrt{a(u +b(x)v)} q \big)&= 0 \,,\notag \\[1ex]
q - \partial_x j(u +b(x)v) &= 0 \,.
\end{align}
Supplemented by some initial condition for $u(t=0,x)$ and $v(t=0,x)$.
Hence the system \labelcref{eq:originalEq} now contains two \textit{instationary} equations for the RG-time evolution of $u$ and $v$ and one additional \textit{stationary} equation for $q$.
The flux of the stationary equation is in a conservative form and given by
\begin{align}\label{eq:source}
	j(s) = \int_0^{s} \sqrt{a(s')} \mathrm{d}s'\, ,
\end{align}
which follows simply from applying the chain rule
\begin{align}
	\sqrt{a(s)} \partial_x s = \partial_s j(s) \partial_x s = \partial_x j(s)\, ,
\end{align}
with $s=u+ b(x) v$.
We note that this particular formulation is insofar different from its original formulation in \cite{Cockburn98thelocal} as $q$ is dependent on two variables through $s$, $u$ and $v$.

\subsection{Numerical fluxes}\label{sec:numflux}

In order to numerically solve the system of conservation equations given in \labelcref{eq:DGForm} over an interval $\Omega$, we introduce a one-dimensional grid over a computational interval $\Omega_h$. $\Omega_h$ is separated into $K$ non-overlapping elements $D^k$, following the notation from \cite{Grossi:2019urj}.
Consider the full solution to be given by ${\bf w} = (u, v, q)^t$, then the local, approximate solution ${\bf w}_h^k = (u_h^k , v_h^k , q_h^k )^t$ is described by a polynomial of degree $P$ in every cell $D^k$,
\begin{align}\label{eq:Approx}
	{\bf w}_h^k (t,x) = \sum_{p=1}^{P+1} {\bf w}_p^k(t) \psi_p(x) \,.
\end{align}
Here we also introduce standard notation from the DG literature for the average $\overline{\bf w}$ and the jump $ [{\bf w}]$ across a cell boundary. Supposing some solution ${\bf w}_h^k$ on the cell $D^k$, we label ${\bf w}^-$ as the solution inside the cell and ${\bf w}^+$ as the solution in the neighbour cell. Then, we define
\begin{align}
  [{\bf w}] = {\bf w}^+ - {\bf w}^- \, , \notag \\[1ex]
  \overline{\bf w} = \frac{{\bf w}^+ + {\bf w}^-}{2}\, .
\end{align}
A solution to \labelcref{eq:DGForm} is only defined in a weak sense, and the resulting system of equations for ${\bf w}_h^k (t,x)$ reads
\begin{widetext}
\begin{align}\notag
	\displaystyle \int_{D^k} \Big( (\partial_t u_h^k )\psi_p - F_h^k( u_h^k ) \partial_x \psi_p \Big) \, \mathrm{d}x
	&= - \int_{\partial D^k} \mathrm{h}_{\mathrm{conv},1} \cdot \hat{n}\, \psi_p  \, \mathrm{d} x \,,  \\[1ex] \notag
	\displaystyle \int_{D^k} \Big( (\partial_t v_h^k )\psi_p -\left(G_h^k( u_h^k , v_h^k ) - \, \sqrt{a_h^k(u_h^k+b_h^k(x) v_h^k)} \, q_h^k \right) \partial_x \psi_p \Big) \, \mathrm{d}x
	&= - \int_{\partial D^k} \left( \mathrm{h}_{\mathrm{conv},2} + \mathrm{h}_{\mathrm{diff},2} \right) \cdot \hat{n}\, \psi_p  \, \mathrm{d} x \,,  \\[1ex]
	\displaystyle \int_{D^k} \Big( q_h^k \, \psi_p + j_h^k(u_h^k+b_h^k(x) v_h^k ) \partial_x \psi_p \Big) \, \mathrm{d}x
	&= - \int_{\partial D^k} \mathrm{h}_{\mathrm{diff},3} \cdot \hat{n}\, \psi_p  \, \mathrm{d} x
\, ,
\label{eq:res}
\end{align}
\end{widetext}
where $\hat{n}$ is the outwards facing unit normal vector at a cell boundary.

The numerical fluxes, at the right-hand side of \labelcref{eq:res} have been separated into convective and diffusive contributions
\begin{align}
  {\bf h}({\bf w^-},{\bf w^+}) = {\bf h}_{\mathrm{conv}}({\bf w^-},{\bf w^+}) +{\bf h}_{\mathrm{diff}}({\bf w^-},{\bf w^+}) \, .
\end{align}
For the convective flux we choose the standard LLF-flux, which has already been used in the direct DG implementations \cite{Grossi:2019urj, Grossi:2021ksl}, given by
\begin{align}\label{eq:LLF}
	{\bf h}_{\mathrm{conv}}({\bf w^-},{\bf w^+}) =
	\begin{pmatrix}
		\bar{F} \\
		\bar{G} \\
		0
	\end{pmatrix}
	- \frac{C_{\mathrm{conv}}}{2} [{\bf w}] \,.
\end{align}
$C_{\mathrm{conv}}$ corresponds to the speed of information propagation across the interface and is chosen as the largest eigenvalue (or an approximation thereof) of the Jacobian
\begin{align}
  J = \begin{pmatrix}
    \partial_{w_1}F & \partial_{w_2}F \\
    \partial_{w_1}G & \partial_{w_2}G
  \end{pmatrix}\, .
\end{align}
It produces an additional diffusive smoothing factor, which enforces the continuity of the solution between cells.

The diffusive numerical flux, which ensures continuity when a diffusive flow is present, is given by
\begin{align} \label{eq:numDiffFlux}
  {\bf h}_{\mathrm{diff}}({\bf w^-},{\bf w^+}) =
  \begin{pmatrix}
    0 \\
    -\frac{[j(u+b(x)v)]}{[u+b(x)v]} \bar{q} \\
    - \bar{j(u+b(x)v)}
  \end{pmatrix}
  - \frac{C_{\mathrm{diff}}}{2} [{\bf w}] \,,
\end{align}
Here a smoothing across cell boundaries is being applied using a diffusive wave-speed, for which
\begin{align}
  C_{\mathrm{diff}} = \begin{pmatrix}
    0 & 0 & 0 \\
    0 & 0 & c \\
    0 & -c & 0 \\
  \end{pmatrix}\,,
\end{align}
is chosen to emulate the LLF-flux \labelcref{eq:LLF}and we use the approximate wave-speed $c=\sqrt{a}$.

\section{DUNE and the dune-FRGDG module}\label{sec:numFrame}

The showcased results have been obtained by using DUNE, the Distributed and Unified Numerics Environment \cite{dune24:16}, which provides a wide range of tools to solve PDEs numerically.
The method detailed in this work has been implemented within the \emph{dune-FRGDG} module, an extension of DUNE specially for solving fRG flows using discontinuous Galerkin methods. A build containing the models showcased in this work can be freely accessed on the web \cite{LDGGit}. For a detailed guide to installation and usage of the framework consult \Cref{app:NumDet}.
Here, we will show very generally how the fRG flows are implemented at the level of the front-end of \emph{dune-FRGDG}.

\subsection{Model definitions}\label{sec:modelDefs}

The DUNE framework, as well as \emph{dune-FRGDG}, use an object-oriented approach to structuring the code. Thus, a system of LDG equations is represented by classes which contain methods and member variables corresponding to the ingredients detailed in \Cref{sec:Numerics}.
The main focus in \emph{dune-FRGDG} is on the model classes, which specify a set of equations to be solved on a computational grid. Their location within the folder structure of the model and functional dependencies are outlined in \Cref{fig:folderStructure}.

When setting up the computation of a specific system of equations, we distinguish between the \textit{instationary} and \textit{stationary} equations, contained in the model classes \texttt{iModel} and \texttt{sModel} respectively. Every model class is derived from a template class, which is contained in the file \texttt{models/modelinterface.hh}. There, every implemented building block of a model is predefined and set to fill the equations with zeros.
This way the user can redefine all methods that are needed for a specific physical model in the derived class, whilst everything else defaults to do nothing.
Therefore, first of all, we need to choose the correct template from which we derive the model class. The signatures of the two model classes thus read
\begin{lstlisting}[frame=single]
class iModel:
	public ModelInterfaceLDGinstat<GV,2,1>;
class sModel:
	public ModelInterfaceLDGstat<GV,2,1>;
\end{lstlisting}
where also the number of components in each class is specified.
\texttt{iModel} contains the flow for the two-component vector \texttt{Range0 u} $ = (u, v)^t$ and \texttt{sModel} specifies the equation for the single component vector \texttt{Range1 q} $ = (q)^t$ - there are two instationary components and one stationary one, corresponding to the LDG system of the O(N) model in \Cref{sec:ON}. We will use the O(N) model from here on as an implementation example. The model can be found in \texttt{models/ON.hh}.
The methods in the model-classes are called for every single element and interface separately. This enables the parallelisation of the code: if the grid is split into several sub-grids these only need to communicate on a few interfaces, since in general every cell only depends directly on its Neighbors due to the numerical fluxes.

Let us focus first on the instationary equations in the \texttt{iModel} class:\\
First, we implement the initial conditions. The initial parameters in the UV are determined in the constructor of the model-class, where they are directly pulled from the initialisation file via \texttt{ptree} (cf. \Cref{app:inidat}):
\begin{lstlisting}[frame=single]
iModel(Dune::ParameterTree ptree_)
: MI(ptree_)
{
  l2 = ptree.get("param.l2", RF(0));
  l4 = ptree.get("param.l4", RF(71.6));
  l8 = ptree.get("param.l8", RF(71.6));
}
\end{lstlisting}
The method \texttt{u0} specifies the initial conditions at the beginning of the simulation, as set in \labelcref{eq:iniC}:
\begin{lstlisting}[frame=single]
Range0 u0(const E &e, const X &x) const
{
  const X xg = e.geometry().global(x);
  Range0 u(0.0);
  u[0] = l2 + xg[0]*l4 + powr<3>(xg[0])*l8;
  u[1] = l4 + 3.*powr<2>(xg[0])*l8;
  return u;
}
\end{lstlisting}
In this example \texttt{e} is a grid parameter, indicating the specific grid cell on which the method is evaluated, whereas \texttt{x} gives the position within the cell. From this we get the global position \texttt{xg}, which is a vector in the dimension of the grid.

To get the code formulation of the flow we insert the ingredients from the physical model \labelcref{eq:ONequations} into the weak formulation of the equations \labelcref{eq:res}, after which ingredient in \labelcref{eq:res} corresponds to a method in the model class. \Cref{sec:NumericsLDG} specified the equations that can be solved within this framework and we will now identify the components of the general mould \labelcref{eq:DGForm} and show their counterpart in the model-classes. The following code-snippets abbreviate the parameter lists of the methods by '...', as they become too lengthy to show here and do not contain important information.
We recapitulate the first line of the general mould \labelcref{eq:DGForm}
\begin{align}
  \partial_t u + \partial_x  F(u, v) &= 0\, .
\end{align}
The DG discretisation of this equation is the first line in \labelcref{eq:res}. The only thing we need here is the definition of the flux $F$ and the numerical flux at the interfaces $F^*$.
\begin{lstlisting}[frame=single]
void flux(...) const
{
  const X xg = e.geometry().global(x);

  const RF F = f<N-1>(u[0])
    + f<1>(u[0] + 2.*xg[0]*u[1]);
  const RF G = u[1] * f<N-1,1>(u[0]);

  Flux[0][0] = F;
  Flux[1][0] = G;
}
\end{lstlisting}
\texttt{RF} is the type of floating point numbers we wish to use - this is aliased to \texttt{double} in the implementation shown here.

The numerical flux is implemented in a separate file; the choice is done within the \emph{Simulation Set} where all general parameters of the simulation are fixed (e.g. the grid, basis functions, etc) - c.f. \Cref{app:numset}.
For specific fluxes one needs additional methods, e.g. for $F^\star$ being an LLF-flux we need to implement a method \texttt{maxeigenvalues} whose signature can be found in the base class.
In the above code snippet, we have also implemented the first part of the second instationary equation, i.e. for the second variable $v$, as it is also discretized in the standard DG way
\begin{align}
	\partial_t v + \partial_x \big(G( u, v) - \sqrt{a(s)}q \big)&= 0
\, .
\end{align}
Next, we need to implement the diffusive component $-\sqrt{a(s)}q$ of the flow of $v$. This is done in the following snippet
\begin{lstlisting}[frame=single]
void diffFlux(...) const
{
  const X xg = e.geometry().global(x);
  const RF s = u[0] + 2.*xg[0]*u[1];

  A[1][0] = -std::sqrt(A_d())*k/(k2 + s) * q[0];
}
\end{lstlisting}
The variable \texttt{A} corresponds to the diffusive component, where the first index of \texttt{A} labels the respective component of ${\bf w}=(u,v)^t$ and the second one the 'spatial' direction of the diffusion. As we only solve equations on one-dimensional intervals in this work, the second index is always $0$.

The numerical flux (i.e. the flow between neighbouring elements) of the diffusive part is implemented in the following.
\begin{lstlisting}[frame=single]
void numericalDiffFlux(...) const
{
  const X xg_s = inside.geometry().
    global(x_inside);
  const X xg_n = outside.geometry().
    global(x_outside);

  const RF s_s = u_s[0] + 2.*xg_s[0]*u_s[1];
  const RF s_n = u_n[0] + 2.*xg_n[0]*u_n[1];
  const RF diffusion = std::sqrt(A_d())*k
    * 0.5*(1/(k2 + s_n) + 1/(k2 + s_s));

  if (!utils::isEqual(s_s, s_n))
    A[1][0] = -std::sqrt(A_d())*k *
      std::log((k2+s_s)/(k2+s_n))/(s_s-s_n) *
      0.5*(q_s[0] + q_n[0]);
  else
    A[1][0] = - diffusion*0.5*(q_s[0] + q_n[0]);
  beta[1][0] = diffusion*(q_n[0] - q_s[0]);
}
\end{lstlisting}
Here, the method is given the information of the inside and outside cell at the border on which it is being called. The first few lines just calculate the quantities at the inside and outside of the cell border. \texttt{A} now plays the role of the first contribution to the diffusive numerical flux in \labelcref{eq:numDiffFlux}, whereas \texttt{beta} implements the additional smoothing in the numerical flux and corresponds to $C_{\mathrm{diff}} [{\bf w}]$. A distinction is made for the case when $s$ is identical on both sides of the cell border, in which case we replace $\frac{[j]}{[s]}$ with the derivative $\frac{\partial j}{\partial s}$.
Since there is no diffusive flux in the first component $u$, the entries at \texttt{A[0][0]} and \texttt{beta[0][0]} remain zero.

Lastly, it remains to implement the stationary equation
\begin{align}
	q - \partial_x j(u+v) &= 0
\, .
\end{align}
This is done in the second model class \texttt{sModel} within the same file. First, we need to implement the flux, analogous to the implementation of the instationary flux
\begin{lstlisting}[frame=single]
void flux(...) const
{
  const X xg = e.geometry().global(x);
  const RF s = u[0] + 2.*xg[0]*u[1];

  if (s > 0)
    F[0][0] = -std::sqrt(A_d())*k
        * std::log(k2 + s);
  else
    F[0][0] = std::sqrt(A_d())*k
        * std::log(1./(k2 + s));
}
\end{lstlisting}
One more thing is missing: The numerical flux of the stationary equation also contains a smoothing term. This is implemented in the following
\begin{lstlisting}[frame=single]
void numericalDiffFlux(...) const
{
  const X xg_s = inside.geometry().
    global(x_inside);
  const X xg_n = outside.geometry().
    global(x_outside);

  const RF s_s = u_s[0] + 2.*xg_s[0]*u_s[1];
  const RF s_n = u_n[0] + 2.*xg_n[0]*u_n[1];

  const RF diffusion = std::sqrt(A_d())*k
    * 0.5*(1/(k2 + s_n) + 1/(k2 + s_s));

  beta[0][0] = - diffusion * (u_n[1] - u_s[1]);
}
\end{lstlisting}
Here, \texttt{beta} corresponds to the stationary, i.e. third, component of $C_{\mathrm{diff}} [{\bf w}]$.

In general, the code is designed to be easily adaptable and can be fitted to any similar problem. For example, adding another instationary equation for a variable to the system is as easy as increasing the first index of the class signature by one and adding in the flow of this third variable in the equations shown here.

\subsection{Simulation Results}

The results of any simulation are saved in VTK-files at times specified by the user (cf. \Cref{app:inidat}). These can be displayed and examined either using the open-source vtk file viewer ParaView, or read into python or Mathematica scripts using openly available vtk libraries. All results in this work have been processed using Mathematica.

\section{Zero dimensional QFT}
\label{sec:HarmO}

We start by investigating the zero dimensional case, which provides a good testing ground for new methods, as it is conceptually rather simple. Furthermore, this limit has recently been thoroughly investigated~\cite{Koenigstein:2021syz, Koenigstein:2021rxj, Steil:2021cbu,Ihssen:2022xjv}. It provides an excellent testing ground for method development, since benchmark results can easily be obtained from a direct evaluation of the generating functional \labelcref{eq:Z-J}. Simply put, it is a simple variant of the O(N)-model, with $N=1$ and $d=0$, where $d$ is the dimension of space-time. The full O(N)-model will be introduced in \Cref{sec:ON}.
Consequently, its very interesting from a computational point of view, since the equations are structurally similar to the full case. Nevertheless, some simplifications arise, which we will detail in the next subsection.

\subsection{Model and flow equation}

The full Quantum Effective Action for a scalar field in zero space-time dimensions is given by the effective potential
\begin{align}
	\Gamma_k[\phi] = V_k(\phi)
\, .
\end{align}

Even in higher dimensions, this simple expansion already captures essential features of the theory at hand as it encodes the coupled RG flow of infinitely many couplings.
Utilizing the Litim cutoff, c.f.~\labelcref{eq:litim_reg}, we can proceed in the usual fashion and obtain
\begin{align}\label{eq:FlowQM}
	\partial_t V_k(\phi) = - \frac{k^2}{k^2 + \partial_\phi^2 V_k(\phi) }
	\,.
\end{align}
Here $k^2 + \partial_\phi^2 V_k(\phi)$ is the regularized propagator, which must be a positive quantity for the equation. In the current case, this property is conserved~\cite{Litim:2006nn}.
Furthermore, with standard initial conditions the resulting effective potential is strictly convex for $k\to0$ even if $m^2 < 0$. This is in contrast to higher dimensions, where the resulting effective potential is still convex but not necessarily strictly convex. This is linked to the absence of spontaneous symmetry breaking in low dimensions. As a consequence, the minimum of the effective potential is $\phi_0 = 0$, while otherwise a non-zero value would be possible. Due to the convexity of effective potential, a flat region can emerge $V_{k=0}(\phi \leq \phi_0) = 0$.
The absence of such a regime simplifies the numerical effort considerably. Nevertheless, discretizing the system proceeds in the same fashion, making low dimensions systems very attractive to benchmark numerical methods in QFT.

\subsection{A simple diffusive system}
\begin{figure*}
	\begin{minipage}[b]{0.4925\linewidth}
		\includegraphics[width=\linewidth]{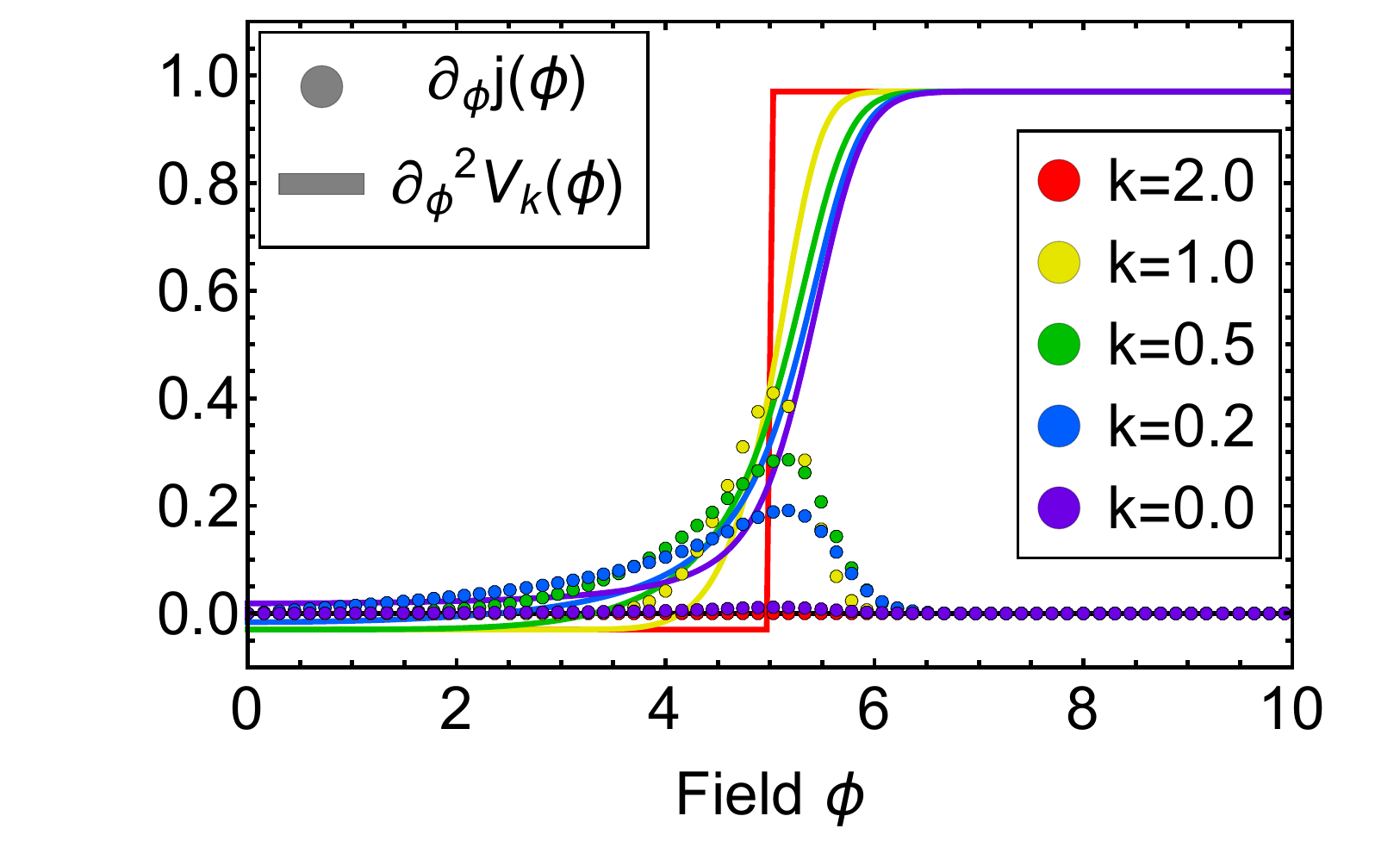}
		\subcaption{Riemann Problem for $v_0 = 0$: The jump smoothes out, but does not travel. The diffusion flow vanishes, while flattening out around the jump.\vspace{4mm} \hspace*{\fill}}
		\label{fig:rieFreeze}
	\end{minipage}%
	\hspace{0.01\linewidth}%
	\begin{minipage}[b]{0.4925\linewidth}
		\includegraphics[width=\linewidth]{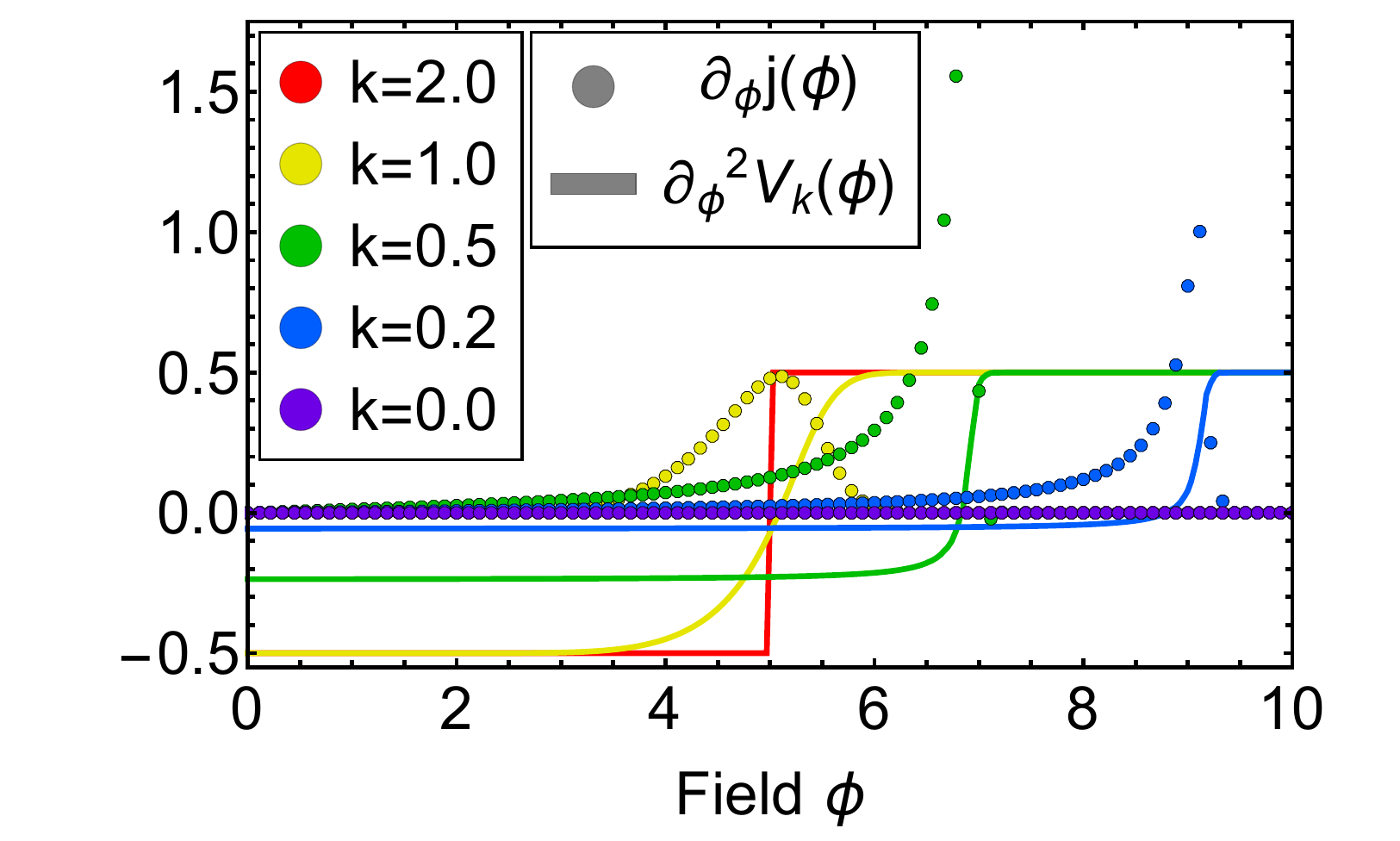}
		\subcaption{Riemann Problem for $v_0 = -0.5$: The jump travels with ongoing RG-time and leaves the grid. The diffusion flow vanishes eventually at $k=0$, but travels shock-like with the jump towards higher grid values. \hspace*{\fill}}
		\label{fig:rieFlow}
	\end{minipage}
	\caption{RG-time evolution of the Riemann problem in $d=0$. We plot the stationary diffusion flux $q=\partial_\phi j(\phi)$ and the second derivative of the potential $\partial_\phi^2 V_k(\phi)$, which are the numerical solutions to the system given in \labelcref{eq:DGflowqm}. \Cref{fig:rieFreeze} shows a freezing in of the jump position, whereas \Cref{fig:rieFlow} shows the jump travelling to higher field values before vanishing. In both cases the sharp jump is flattened out, as expected of a diffusive system. \hspace*{\fill}}
	\label{fig:rie}
\end{figure*}

To apply the LDG method from \Cref{sec:Numerics} to the zero dimensional case, the equation must be reformulated such that it is a closed expression of the second derivative of the potential, which we rename $v(\phi) =\partial_\phi^2 V_k(\phi)$. Similarly to the procedure in \cite{Grossi:2019urj,Grossi:2021ksl}, we take two derivatives with respect to $\phi$ of the flow\labelcref{eq:FlowQM} to close the formulation in $v(\phi) = \partial_\phi^2 V_k(\phi)$. All information generated by the flow is contained within the second derivative. This can already be inferred from the fact that the fundamental object in functional formulations is the propagator and therefore the two-point function, which is by definition a (functional) derivative. Therefore, we introduce the schematic flow equation
\begin{align}
\label{eq:DGflowqm}
	\partial_t v = \partial_\phi^2 f(v) = \partial_\phi \big( \partial_v f(v) \  \partial_\phi v \big)
\, .
\end{align}
We can now make use of the formalism given in \labelcref{eq:DGForm} and indicate the  corresponding functions in the code. Since we are considering a one component system we set $u = 0$ and thus the convective flux vanishes, i.e.~$F(u,v)=0$. The code reads \texttt{ModelInterfaceLDGinstat<G,1,1>} and \texttt{ModelInterfaceLDGstat<GV,1,1>}. In the equation for $v(\phi)$, the convective contribution $G(u,v)$ (\texttt{flux()} in \texttt{iModel}) is also set to zero, since we are considering a purely diffusive system. Now, let us consider the diffusive flux (\texttt{diffFlux()} in \texttt{iModel}) where we have, $s = v$, c.f. the discussion around \labelcref{eq:source}, as well as
\begin{align}
	\sqrt{a(v)} &= \frac{k}{k^2+v}\, .
\end{align}
By performing the integration we obtain the stationary flux (\texttt{flux()} in \texttt{sModel})
\begin{align}
\label{eq:zero_diff_flux}
	j(v)=& k \log(k^2+v)
\, .
\end{align}
Please note, that we dropped a constant term in \labelcref{eq:zero_diff_flux}, which would arise from the integration boundaries in the original definition \labelcref{eq:source}. This term does not contribute here because the flow $f(v)$ in \labelcref{eq:DGflowqm} does not explicitly depend on $\phi$.

\subsection{Results}

\subsubsection{The Riemann Problem}

We study the behaviour of this simple diffusive system in a first order regime. This is not only a common check for how the scheme deals with jumps in the solution, but is also interesting for physical systems in critical regions. For example, it was found in \cite{Grossi:2021ksl} that the quark-meson model in the large-$N$ limit displays shock development at high densities $\mu$. With the following investigation we can now expand on how an addition of diffusion would affect these systems.

We consider initial conditions that contain a single jump in the solution. This is also known as the Riemann-problem
\begin{align}
	v(\phi)\vert_{t=0} = v_0 + \theta (\phi-5) A \,,
\end{align}
where the jump is situated in the middle of the numerical grid $\phi \in \{0,10 \}$, $v_0$ is the initial value in the regime $\phi\leq5$ and $|A|=1$ the height of the jump. All computations start at an RG-scale of $\Lambda=2$. For clarification on implementation and the numerical intricacies see \Cref{sec:numFrame} and \Cref{app:NumDet}.

\Cref{fig:rie} shows that the diffusive contribution to the flux may or may not travel with the RG-time, depending on the initial value $v_0$. \Cref{fig:rieFreeze} shows a scenario with little dynamics: The diffusion is smoothing out the jump until it vanishes with $k \to 0$. Throughout this process the diffusion-peak-position, and therefore also the position of the steepest slope in the potential derivative, are barely moving.
In contrast to this, \Cref{fig:rieFlow} shows the dynamical case where the diffusion flux is first smoothing out the jump and subsequently pushing the original jump to higher values of the spatial coordinate $\phi$.
In this dynamical scenario the diffusive flux is developing a discontinuity at intermediate RG-times, which eventually vanishes again in the limit $k\to 0$.

The result of this simple consideration is already remarkable. The jumps do not get flattened away like in usual diffusion equations. The width freezes to a finite value, demonstrating support for almost shock like structures throughout the RG flow. In comparison to ordinary dispersive shock waves, no ringing arises in the vicinity of the jump.

\subsubsection{Non-Analytic 1st derivative}
\begin{figure*}
	\begin{minipage}[b]{0.4925\linewidth}
		\includegraphics[width=\linewidth]{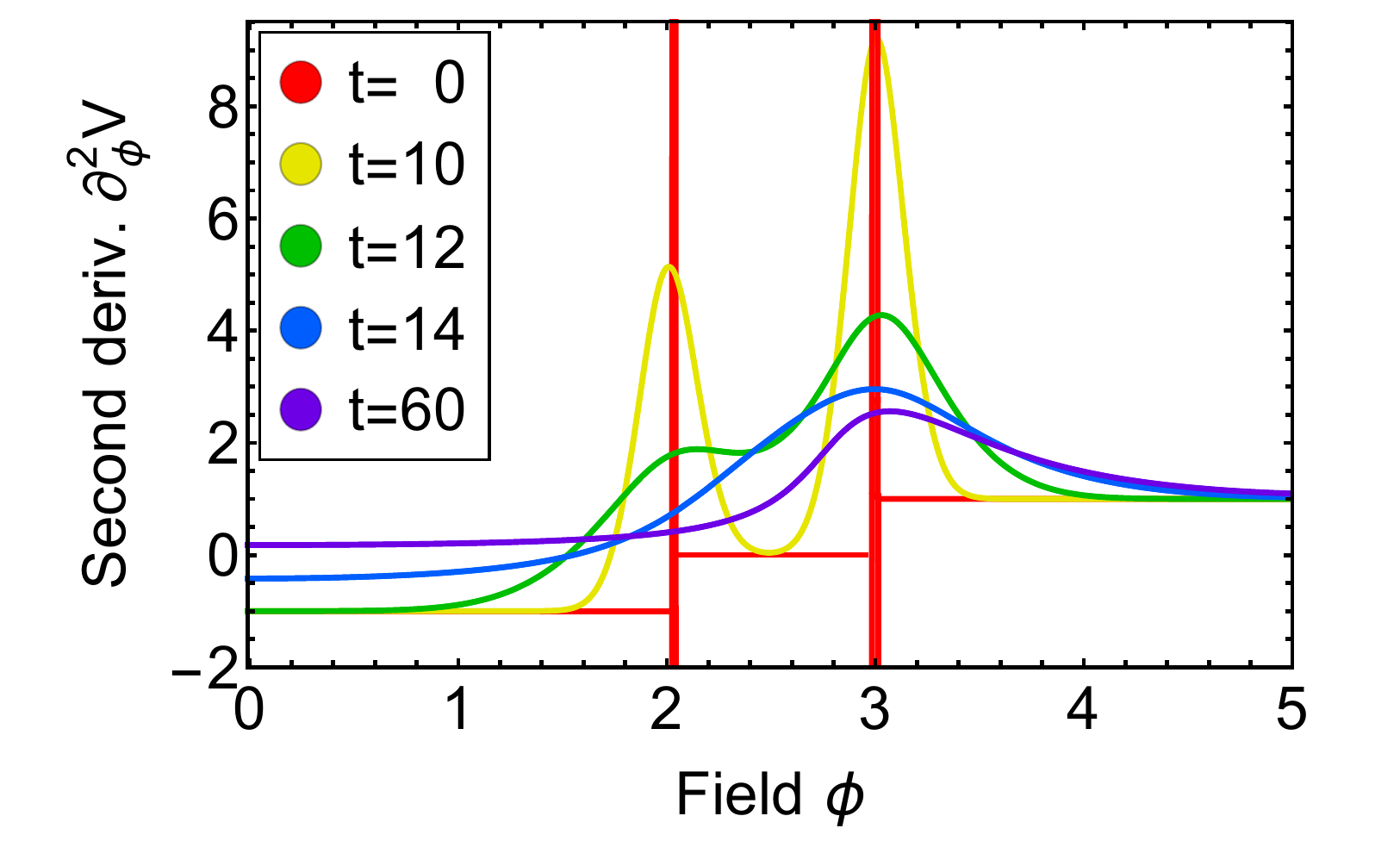}
		\subcaption{Numerical solution to the flow of the initial conditions given in \labelcref{eq:disc_zero_pot}, in terms of the second derivative $v(\phi) = \partial_\phi^2 V(\phi)$. The singularity in the initial conditions is projected onto the polynomial basis \labelcref{eq:polybasis}.\hspace*{\fill}}
		\label{fig:comp_zero_deriv}
	\end{minipage}%
	\hspace{0.01\linewidth}%
	\begin{minipage}[b]{0.4925\linewidth}
		\includegraphics[width=\linewidth]{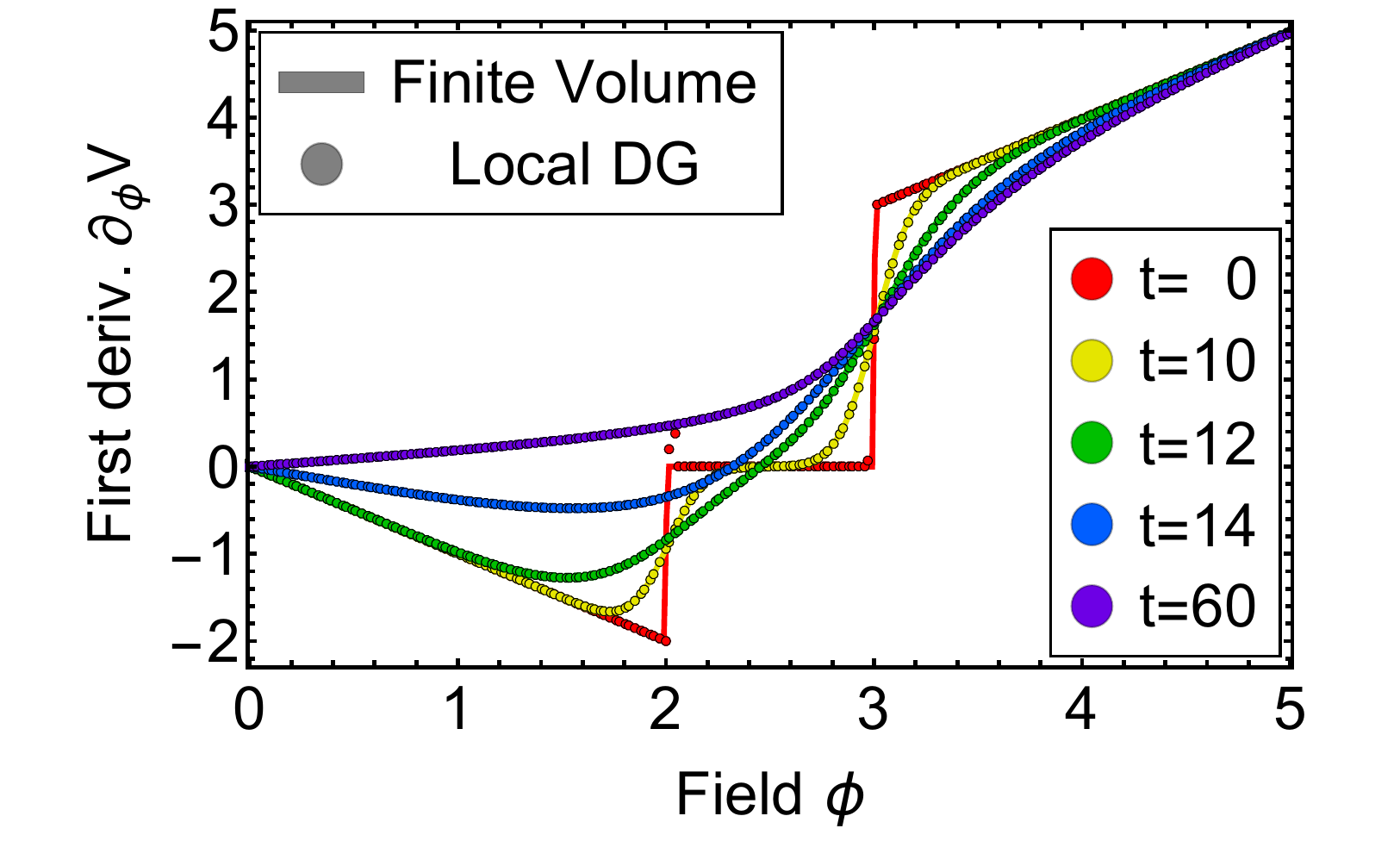}
		\subcaption{Comparison of the flow of the first derivative of the initial conditions given in \labelcref{eq:disc_zero_pot}. The LDG result is obtained by integrating the numerical solution. The finite volume result is taken from \cite{Koenigstein:2021rxj}. \hspace*{\fill}}
		\label{fig:comp_zero_func}
	\end{minipage}
	\caption{RG-time dependence of the potential in \labelcref{eq:disc_zero_pot} in terms of its derivatives.\hspace*{\fill}}
	\label{fig:comp_zero}
\end{figure*}
In the spirit of \cite{Koenigstein:2021rxj} (Sec. V.A: Test case I), we study a UV-potential with kinks. 
The initial potential is given by
\begin{align} \label{eq:disc_zero_pot}
	V(\phi) = 
	\begin{cases}
		-\frac{1}{2} \phi^2\,, & \phi \leq 2 \,, \\[1ex]
		-2 \,, & 2 < \phi \leq 2  \,, \\[1ex]
		\frac{1}{2} (\phi^2 - 13) \,, & 3 < \phi  \,,
	\end{cases}
\end{align}
at an initial UV-cutoff scale $\Lambda = 10^6$. In order to match the calculation in \cite{Koenigstein:2021rxj} we adapt the same regulator $R_k(p) = \Lambda e^{-t}$, which modifies the flow equations only slightly. Previous studies \cite{Grossi:2019urj,Grossi:2021ksl, Koenigstein:2021rxj, Koenigstein:2021syz} focus on solving fRG flow equations in terms of the first potential derivative $u = \partial_\phi V$. Hence, \labelcref{eq:disc_zero_pot} was designed with the Riemann problem in mind, since kinks in the potential translate to a discontinuity in $u$. The LDG framework is formulated in terms of the second derivative $v = \partial_\phi u$, which requires us to resolve delta-peaks at $|\phi|=2, \ 3$. In the weak formulation \labelcref{eq:res} this amounts to a projection of the initial data onto the polynomial basis \labelcref{eq:polybasis} at hand.

The solution to the flow is shown in \Cref{fig:comp_zero}. The computational domain $\Omega = [-6,6]$ is divided in $K=101$ equally sized cells. The chosen value for $K$ ensures that the initial singularities at $\Lambda$ are not situated on cell interfaces. Within cells, we use a polynomial order of $P=4$. This choice demonstrates nicely, how the flux-formulation of the LDG scheme is able to deal with oscillatory behaviour around discontinuities. We compare our solution to the calculation of \cite{Koenigstein:2021rxj}, shown in \Cref{fig:comp_zero_func}, where this initial potential was first considered. The results are nicely in agreement, as expected.
\Cref{fig:comp_zero_deriv} shows the direct numerical solution of \labelcref{eq:FlowQM} in terms of the second derivative $v = \partial_\phi u$. Here, oscillations appear at early RG-times due to the projection of delta-peaks on a polynomial basis.

\section{O(N) model}

\label{sec:ON}
We now turn our attention to the O(N)-model. It is a common test subject when first considering new methods due to its simplicity. Simultaneously, it has a large range of applications, ranging from condensed matter physics over cosmology to QCD, see~\cite{Dupuis:2020fhh} and references therein.

Within the fRG, it has been studied extensively~\cite{Berges:1996ib, Papp:1999he, Litim:2001up, Litim:2001dt, Litim:2003kf, Blaizot:2005xy, Litim:2005us, Schaefer:2006ds, Pelaez:2015nsa, Borchardt:2015rxa, Borchardt:2016pif, Litim:2016hlb, Juttner:2017cpr, Litim:2017cnl, Litim:2018pxe, Balog:2019rrg,Grossi:2019urj, DePolsi:2020pjk, Koenigstein:2021syz, Koenigstein:2021rxj, Steil:2021cbu, DePolsi:2021cmi, DePolsi:2022wyb}, ranging from fixed points, to its RG-time evolution in physical systems. Therefore, it constitutes an excellent toy-model for our purpose. Particularly, it features second order phase transitions and allows, in principle, for the possibility of first-order phase transitions.
%

\subsection{Model and flow equation}

The average effective action of the O(N)-model reads at first order in the derivative expansion
\begin{equation}\label{eq:effectiveAction}
	\Gamma_k[\phi] = \int d^{d}x\, \bigg\{\frac{1}{2} Z_{k}(\rho)(\partial_\mu\phi)^2 + V_k(\rho)-c_\sigma \sigma \bigg\}\,,
\end{equation}
where $\phi = (\sigma, \vec{\pi})^t$ and the potential $V$ only depends on the O(N) invariant $\rho=\frac{\sigma^2 + \vec{\pi}^2}{2}$. This notation follows the usual applications of the O(N) model to QCD. Hereby, we parametrize the ground-state without loss of generality as $\phi_0 = (\sigma, \vec{0})$. The factor $-c_\sigma \sigma$ corresponds to the explicit symmetry breaking induced by bare quark masses and decouples from the RG flow, since it is a term linear in the field. It only enters when solving the equation of motion after performing the RG integration and can therefore be ignored in the following.

The flow equation is derived from the Wetterich equation \labelcref{eq:wetterich}, by evaluating the scalar field at $\phi_0$. For simplicity we restrict ourselves to Local Potential Approximation (LPA) and set $Z_{k}(\rho) = 1$. The resulting flow is then given by
\begin{align}
\label{eq:FlowEqFiniteN}
	\partial_t V_k(\rho)=
	-\frac{v_d k^{d+2}}{2(2\pi)^d} \Bigg[\frac{N-1}{k^2+m_\pi^2}  +\frac{1}{k^2+m_\sigma^2} \Bigg]
\, ,
\end{align}
where the angular integration gives rise to the factor $v_d = \frac{2 \pi^{d/2}}{\Gamma(d/2) d}$. The curvature masses are given by
\begin{align}
\label{eq:masses}
	m^2_\sigma &= \partial_\rho V_k(\rho) + 2 \rho\, \partial_\rho^2 V_k(\rho) \, , \notag \\
	m^2_\pi &= \partial_\rho V_k(\rho)
\,.
\end{align}
%

\subsection{A convection-diffusion system}
\label{sec:NumericsLDG}

\Cref{eq:FlowEqFiniteN} is reformulated in terms of $u = \partial_\rho V$ and $v = \partial^2_\rho V$. The original flow equation is rewritten as
\begin{align}
	\partial_t V_k(\rho) = f_{N-1}(u) + f_1(u+2 \rho v) \,,
\end{align}
where $f_i(x)=A_d \frac{ i }{1+ \frac{x}{k^2}}$ and $A_d=\frac{v_d k^{d}}{2(2\pi)^d}\,$. This equation does not dependent on the original function $V_k(\rho)$ but only on its first and second $\rho$-derivative. To disentangle this second order partial differential equation we take derivatives of the equation. The derivation is given in \Cref{app:DGFlow}, here we only identify the corresponding terms from \labelcref{eq:DGForm} for an easy application to the numerical framework
\begin{align}\label{eq:ONequations}
	F(u,v) &=   f_{N-1}(u) + f_1(u+2 \rho v)\, , \notag\\[1ex]
	G(u,v) &=   v \ \partial_u f_{N-1}(u)\, ,  \notag\\[1ex]
	\sqrt{a(u+ 2 \rho v)} &=   \frac{A_d^{1/2} k}{k^2+u+ 2 \rho v}\,.
\end{align}
Therefore after integration and following \labelcref{eq:source} we obtain
\begin{align}
\label{eq:ONj}
	j(s)= A_d^{1/2} k \, \log(k^2+s)
	\,,
\end{align}
where $s=u+ 2 \rho v$. As before, we ignored irrelevant constants arising from the integration boundary in \labelcref{eq:ONj}.

\subsection{Results}
\begin{figure}
	\includegraphics[width=\linewidth]{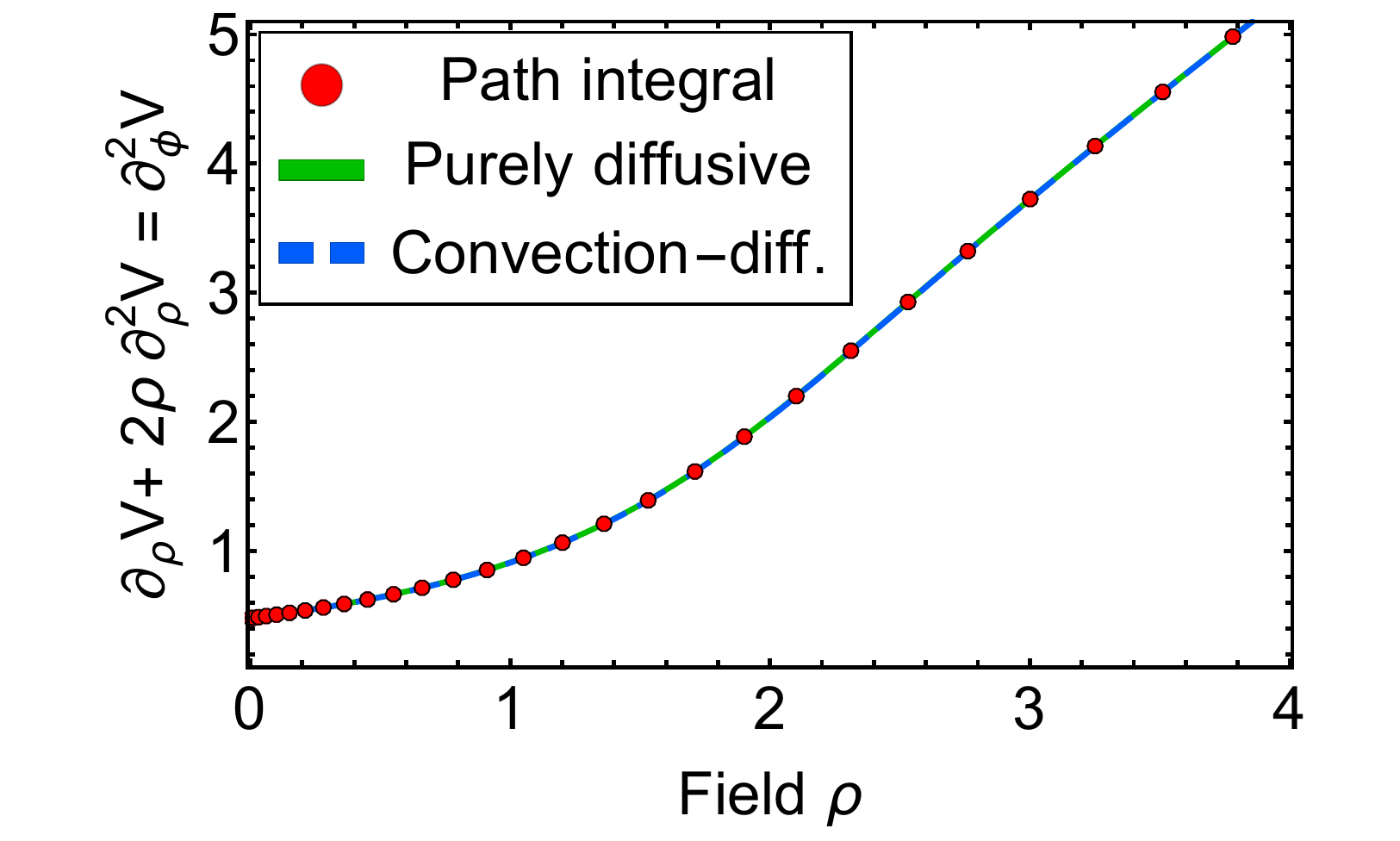}
	\caption{Graphical comparison of the second derivative of the effective potential $V$ as a function of $\rho$. We compare data from a numerical evaluation of the path integral \labelcref{eq:Z-J}, the purely diffusive system \Cref{sec:HarmO} and the convective diffusive system \Cref{sec:ON}. \hspace*{\fill}}
	\label{fig:derivOS}
\end{figure}
In this Section, we perform some benchmark checks. For this purpose we compare to the zero dimensional example from \Cref{sec:HarmO}, as well as the large-$N$ limit, which has been investigated in various contexts in \cite{Grossi:2019urj,Grossi:2021ksl, Steil:2021cbu}.
Both examples test different features of the presented scheme. The zero dimensional case, formulated in terms of the field $\phi$, is a diffusive system. Recovering the results from \Cref{sec:HarmO} therefore checks convergence of the convection-diffusion system in the diffusion dominated scenario. On the other hand, the large-$N$ limit is a purely convective system, since the second derivative term (i.e. the diffusion term) drops out in the limit $N \to \infty$. Performing computations at a very large $N$ therefore allows to check convergence features in a convection dominated scenario. However, at this point we would like to remark that the solution of the equation for $N\to\infty$ does not necessarily coincide with the solution at $N=\infty$, see \cite{Yabunaka:2017uox, Yabunaka:2018mju, Yabunaka:2021fow, Steil:2021cbu} for a critical discussion.

For computations we use a simple $\phi^4$ potential as initial condition, to wit
\begin{align}\label{eq:iniC}
	V_{\Lambda}(\phi) &=  m^2 \rho + \frac{1}{2}\lambda \rho^2 \,.
\end{align}
We chose $m^2=-0.5$, so that we recover a symmetry breaking scenario in $d>2$, as well as $\lambda=0.5$.

\subsubsection{Comparison to the purely diffusive system}

\begin{figure}
	\includegraphics[width=\linewidth]{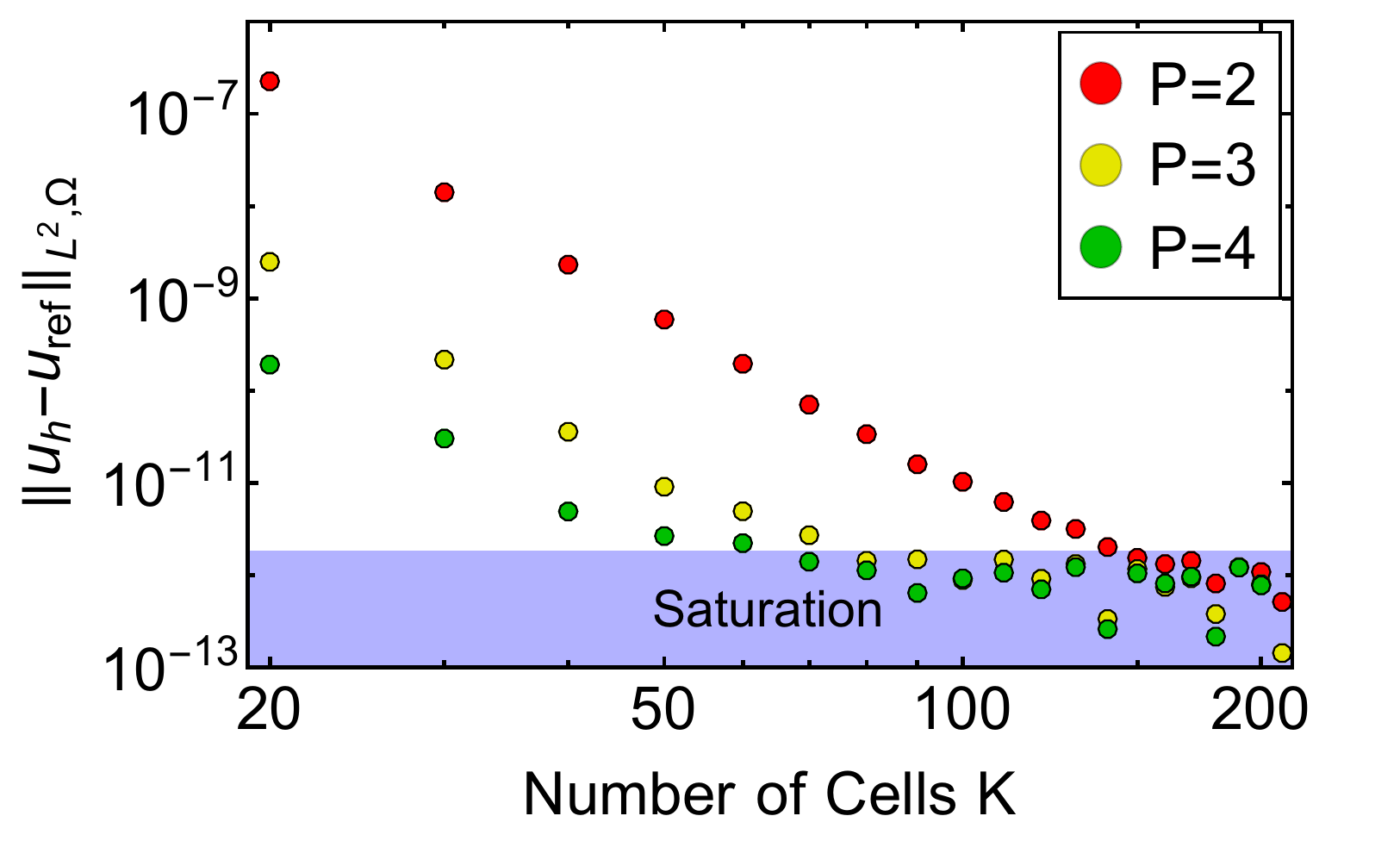}
	\caption{$L^2$ error with respect to a reference computation $u_\text{ref}$. The reference computation is given by the solution with $K=250$ and $P=4$, whereas $u_h$ is the numerical solution for the O(N) computation using different polynomial orders and numbers of cells $K$. \hspace*{\fill}}
	\label{fig:convOS}
\end{figure}
\begin{figure*}
	\begin{minipage}[b]{0.4925\linewidth}
		\includegraphics[width=\linewidth]{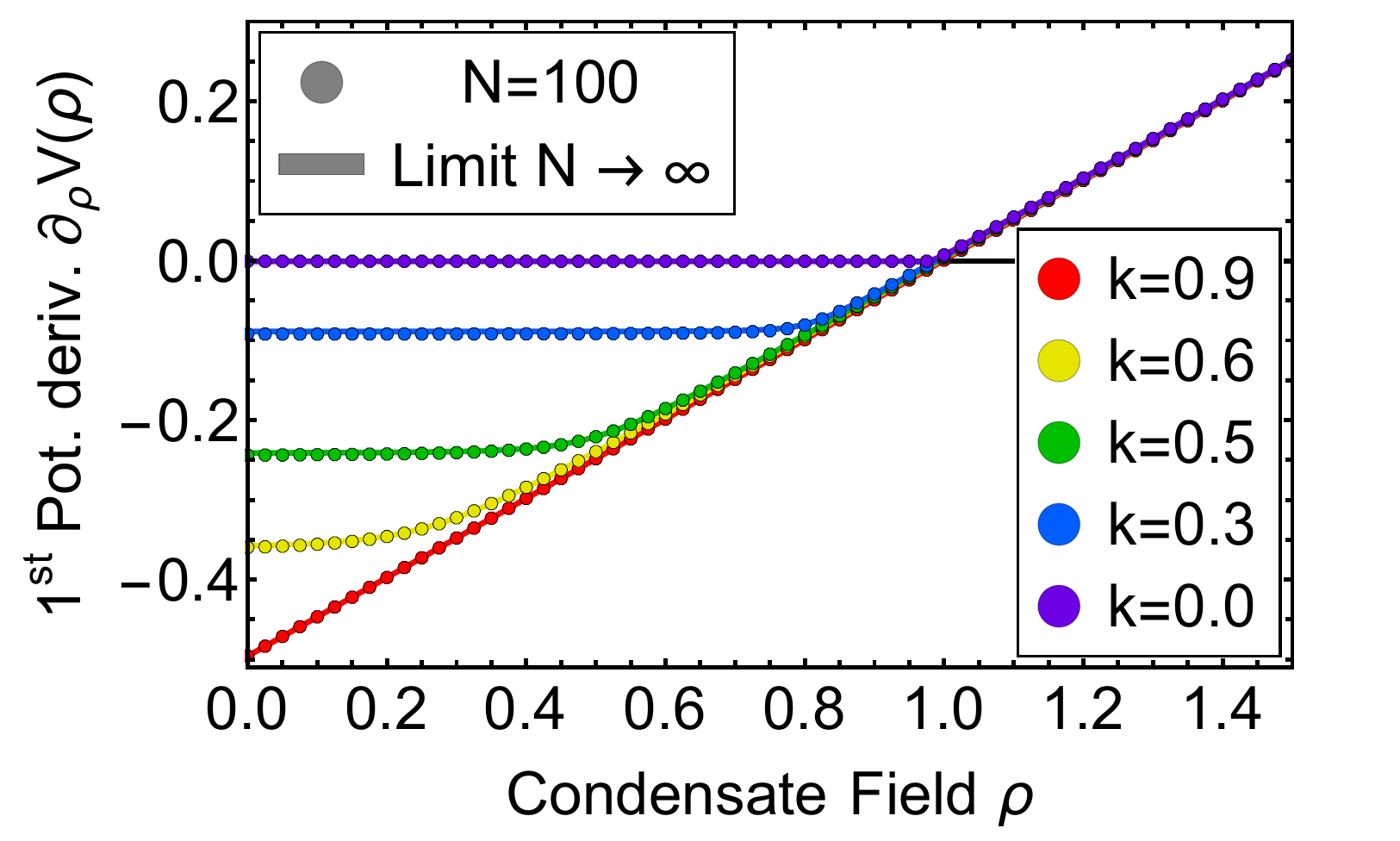}
		\subcaption{Graphical comparison of the first derivative of the effective potential $u = \partial_\rho V$. \hspace*{\fill}}
		\label{fig:convON}
	\end{minipage}%
	\hspace{0.01\linewidth}%
	\begin{minipage}[b]{0.4925\linewidth}
		\includegraphics[width=\linewidth]{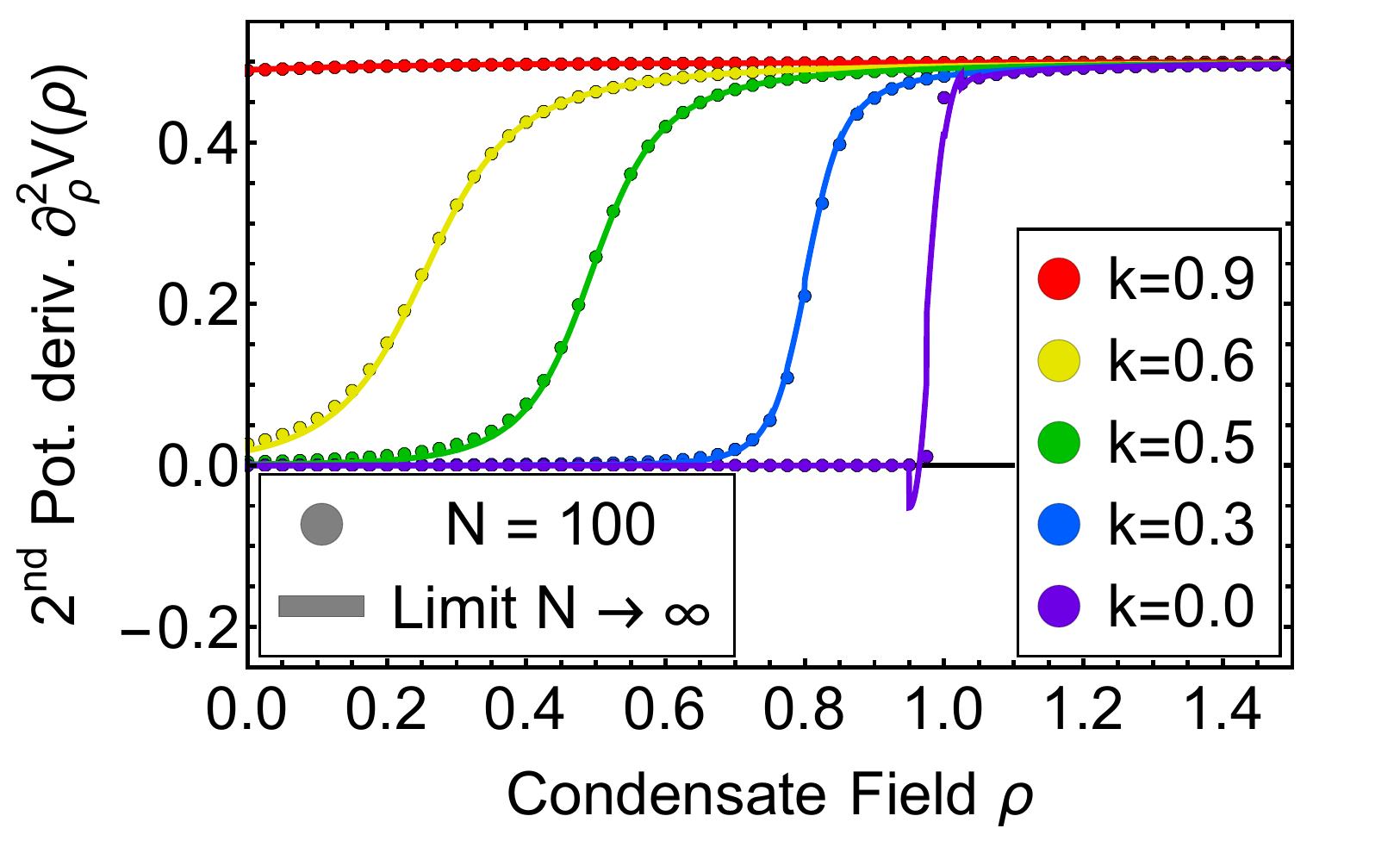}
		\subcaption{Graphical comparison of the second derivative of the effective potential $v = \partial_\rho u = \partial_\rho^2 V$ \labelcref{eq:dxu}. \hspace*{\fill}}
		\label{fig:derivON}
		
	\end{minipage}
	\caption{Graphical comparison of the purely convective large-$N$ limit and the convection-diffusion system \labelcref{eq:FlowEqFiniteN} at $N=100$. We consider the broken symmetry phase in $d=3$ using the initial conditions from \labelcref{eq:iniC}. In the large-$N$ limit, the derivative is obtained by interpolation of the numerical data. This interpolation struggles with the resolution of the kink within the first derivative in (b). \hspace*{\fill}}
	\label{fig:compON}
\end{figure*}
Let us preface the comparison to the purely diffusive scenario with a general discussion of the $O(1)$ case.
In many applications, the O(N) model uses a formulation in the chiral invariant $\rho = \frac{\phi^2}{2}$. This choice of variable imposes the $Z_2$ symmetry of the potential and raises the question of which boundary conditions to use at $\rho \to 0$. By comparison, the formulation in \Cref{sec:HarmO} is given in terms of $\phi$. Here, the symmetry of the potential needs to be checked explicitly. For a throughout investigation of the coordinate choice see \cite{Koenigstein:2021syz}.\\
In particular for $N=1$ the formulation in $\rho$ \labelcref{eq:ONequations} is not ideal. In dimensions $d>2$, where we have a spontaneous symmetry breaking, the flow is strongly governed by the convexity restoring property. This property requires $\frac{u + 2 \rho v}{k^2}>-1$ for the massive mode ($N=1$) and additionally $\frac{u}{k^2}>-1$ for the massless modes ($N>1$), such that the pole in the flow equation \labelcref{eq:ONequations} is not reached at finite $k>0$. This requirement demands increasingly high numerical precision as $k \to 0$ in the broken phase and presents a big challenge for all numerical schemes which attempt to resolve the effective potential in the broken phase, see e.g.~\cite{Pelaez:2015nsa, Grossi:2019urj}. In the standard example of $\phi^4$ initial conditions, convexity restoration is driven by the massless modes. At $N=1$ or for more intricate initial conditions, where $\partial_\rho u < 0$ for some $\rho$, the dynamics are governed by the massive mode such that the relation $\frac{u + 2 \rho v}{k^2}>-1$ has to be fulfilled. While this property is, in principle, enforced by the flow, it requires precision in two distinct expressions $u, \ v$ as opposed to one in the purely diffusive case \labelcref{eq:FlowQM}. A prominent example for the scenarios with $d>2$ and $\partial_\rho u < 0$ is the high density regime of the quark-meson model, or more generally, of QCD. These ingredients to the flow equations cause strong dynamics and potentially shock development \cite{Grossi:2021ksl}. Numerical intricacies connected to shock development and convexity restoration are discussed in \Cref{app:shock}. 

In this sense, a comparison to \Cref{sec:HarmO} is somewhat trivial, since it is simply an O(N) model with $N=1$ and $d=0$. There is no spontaneous symmetry breaking process and the result for a simple $\phi^4$ theory can also be obtained from a numerical evaluation of the path integral \labelcref{eq:Z-J}. Thus, the full effective potential $V_{k=0}(\rho)$ is computed in $d=0$ by a direct numerical evaluation of $Z[J]$ with the classical action specified by the initial conditions in \labelcref{eq:iniC}, following the steps of the derivation in \Cref{sec:FRG}. In order to capture the full result with the fRG, the computation is started at an initial cutoff scale of $\Lambda = 20$, where the numerical result of the modified path integral $Z_k[J]$ \labelcref{eq:modGenFunc} corresponds to the initial conditions $V_{\Lambda}(\rho)$ \labelcref{eq:iniC}.

We now perform a benchmark check between the numerical path integral result, the purely diffusive system in \Cref{sec:HarmO}, and the current convection-diffusion driven system. A graphical comparison is given in \Cref{fig:derivOS}.
To further showcase the convergence of the convective-diffusive LDG scheme we compute a reference solution $u_\mathrm{ref}(\rho)$. This computation uses a grid with $K=250$ cells and a polynomial order $N=4$ for values of $\rho \in [0,8]$. The integration is performed up to an RG-scale $k=0.002$.
This reference solution is compared to numerical solutions $u_h$ using varying polynomial orders $P=2,\ 3, \ 4$ and cell numbers $K \in [20,220]$. For the quantitative comparison we make use of the $L^2$ norm
\begin{equation*}
	\| u_h - u_{\mathrm{ref}} \|_{L^2, \Omega} = \int_{0}^{3.5} \Big( u_h (\rho) - u_\mathrm{ref}(\rho)\Big)^2 d\rho \,.
\end{equation*}
Results are given in \Cref{fig:convOS}. We do not see the expected linear behaviour from \cite{Grossi:2019urj}, since the error saturates at $10^{-12}$ due to the finite floating point precision of the calculation.

\subsubsection{Comparison to the Large-N limit}

In a second benchmark test we recover the large-$N$ limit in the broken symmetry phase. In the limit $N \to \infty$ the massive mode $m_\sigma = u + 2 v \rho$ drops out of the flow completely, resulting in a purely convective system. Previous work investigated the large-$N$ limit from this perspective \cite{Grossi:2019urj,Grossi:2021ksl}.
Using the initial conditions in \labelcref{eq:iniC}, we compute in $d=3$ and $N=100$. This computation is compared to the results from \cite{Grossi:2019urj} in \Cref{fig:compON}, which confirms that the large-$N$ limit is recovered by comparing the first potential derivative $ u = \partial_\rho V$. Furthermore, the LDG scheme allows to precisely resolve the second derivative of the potential $v = \partial_\rho u = \partial_\rho^2 V$ \labelcref{eq:dxu}, see \Cref{fig:derivON}. The feed back of $v$ to the flow of $u$ is suppressed by a factor of $1/99$. Still, $v$ is computed separately and provided with a numerical flux which allows to get rid of numerical oscillations around the clearly visible jump in the limit $k \to 0$ at $\rho \approx 1.0$. 

\Cref{fig:compON} demonstrates the limitations of the DDG scheme at finite $N$. Even though DDG is able to provide higher derivative terms due to the polynomial basis \labelcref{eq:polybasis}, the formulation lacks a numerical flux for explicit higher derivative operators. If the solution $u$ is flat, the missing numerical flux to the finite $N$ flow \labelcref{eq:FlowEqFiniteN} is negligible and computations retain their convergence properties \cite{Grossi:2021ksl}. However, in case of non analyticities, the numerical flux of higher derivative operators can no longer be neglected, which becomes apparent from the oscillations around the jump in the large-$N$ computation for $k \to 0$.

\section{LDG in-depth and perspective}
\label{sec:perspective}

The formulation of the LDG method presented here, which has been adapted from \cite{Cockburn98thelocal}, is of course only one of many.
For example, in the case of higher-dimensional systems, replacing $q - \partial_x j(s) = 0$ with $q - \partial_x s = 0$ and using an alternating flux definition may be desired for a lower performance impact and simplification of the implementation, as has been already noted in \cite{Cockburn98thelocal}. Furthermore, it should be noted that usually the integration in the definition of $j(s)$ can not be performed analytically and thus requires numerical integration.

Other LDG methods may also be desirable, e.g. methods that directly infer $\nabla {\bf w}$ by giving a numerically sensible definition of this quantity, see e.g. \cite{FengLewisHighOrder}, thus making the rewriting into $u$ and $v$ we had to perform in \Cref{sec:Numerics} superfluous. This of course has its drawbacks, as the stability of the method may become suboptimal and requires further method development.

Additionally, fRG flows seem to be strongly dominated by certain substructures of the equations. The regulator derivative in the equations forces all changes to be in a narrow region in momentum space. This property is translated directly to field space, with the exception of the convexity restoring regime, resulting in the convection dominated nature of the equations. One direct consequence is the large asymmetry of the diffusive flux, which lead to a breakdown of the scheme in \cite{Grossi:2021ksl}. This reflects simply the fact that the RG flow evolves from the UV to the IR and information gets propagated along these trajectories,~c.f.~\cite{Grossi:2019urj, Koenigstein:2021rxj} for an extensive discussion. LDG resolves this elegantly, by taking precisely this structural property of the flow into account. From a physics perspective, this is not the end of the line for potentially exploitable structure within these equation. One prominent example is the convexity restoring regime, where the solution approaches zero for a finite domain. In the present equations, its this is driven by a singularity in the analytically continued equation \cite{Litim:2006nn}. Computationally, this is tightly linked to time-stepping.

Not very surprisingly, for more complex and computationally intensive problems a different time-stepping scheme must be considered. Due to convexity restoration when symmetry breaking occurs, the fRG flows become stiff and explicit time stepping becomes highly inefficient, see \Cref{app:shock}. A possible remedy for that is to move to implicit time-stepping schemes or to adapt exponential integrators to the structure of the equations in this regime, see~e.g.~\cite{hochbruck2010exponential} for an introduction.

\section{Summary and outlook}
\label{sec:sum}

In this paper we have applied the Local Discontinuous Galerkin scheme to flow equations in the O(N)-model. This takes the convection dominated structure of the underlying flow equations properly into account and provides an highly efficient scheme. We have applied this scheme to several combinations of the space-time dimension $d$ and the number of field components $N$. Thereby, we benchmarked both, a convection dominated ($N\gg1$) and a purely diffusive ($N=1$) case. 

A substantial part of the paper is dedicated to detailing the numerical implementation, which is made publicly available. Said implementation is a necessary ingredient in our quest to achieve qualitatively and quantitatively reliable results in the phase structure of generic theories. The respective phenomena range from physics at the highest momentum scales with asymptotically safe gravity to condensed matter systems and statistical physics at low momentum scales. 

A particular interest is the resolution of the phase structure of QCD with the functional Renormalisation Group, where a fully reliable access to finite densities indeed requires the resolution of the  two-fold challenge discussed in the introduction. We hope to report on this matter in the near future.

\begin{acknowledgments}
We thank Eduardo Grossi, Nicolas Hendricks, Adrian Koenigstein, {\'A}lvaro Pastor-Guti{\'e}rrez and Valentin W{\"u}st for discussions. This work is done within the fQCD collaboration \cite{fQCD}, and is supported by the Studienstiftung des Deutschen Volkes and EMMI. This work is funded by the Deutsche Forschungsgemeinschaft (DFG, German Research Foundation) under Germany’s Excellence Strategy EXC 2181/1 - 390900948 (the Heidelberg STRUCTURES Excellence Cluster) and the Collaborative Research Centre SFB 1225 (ISOQUANT). NW acknowledges support by the Deutsche Forschungsgemeinschaft (DFG, German Research Foundation) – Project number 315477589 – TRR 211 and by the State of Hesse within the Research Cluster ELEMENTS (Project No. 500/10.006). \\[2ex]
\end{acknowledgments}

\appendix

\section{The dune-FRGDG module}\label{app:NumDet}

We presented a numerical framework to solve equations from the fRG. In particular, we gave details on how to translate flow equations into well tested numerical operators in \Cref{sec:numFrame}.  Now, we focus on the purely numerical details of the implementation and set-up of the freely available \emph{dune-FRGDG} module.
Firstly, \Cref{app:installing} contains detailed instructions on the installation of the framework.
\Cref{app:outline} outlines the general structure of the module. This is followed by a discussion of different components:
In \Cref{app:numset} we explain the \emph{simulation set}, which allows to specify the general numerical settings of the calculation, such as the numerical grid or the choice of numerical fluxes.
The implementation of physical models within the framework has already been discussed in \Cref{sec:numFrame}, but we give some additional remarks on the parameters that can be changed with the \texttt{.ini} files after compiling the programs in \Cref{app:inidat}.

\subsection{Installation}\label{app:installing}

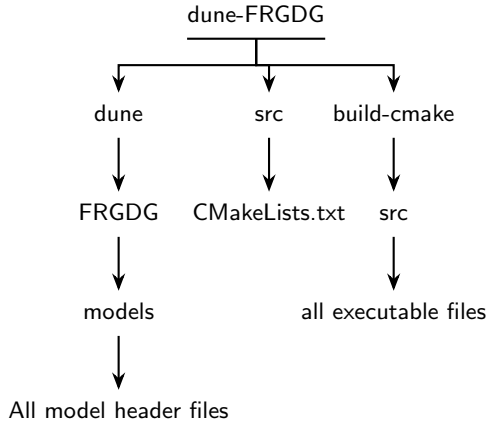
\begin{figure}[h]
	\centering
	\vspace{0.6cm}
	\begin{forest}
		for tree={
			align=center,
			font=\sffamily,
			edge+={thick, -{Stealth[]}},
			l sep'+=10pt,
			fork sep'=10pt,
		},
		forked edges,
		if level=0{
			inner xsep=0pt,
			tikz={\draw [thick] (.children first) -- (.children last);}
		}{},
		[dune-FRGDG
		[dune
		[FRGDG
		[models
		[All model header files]
		]
		]
		]
		[src
		[CMakeLists.txt]
		]
		[build-cmake
		[src
		[all executable files]
		]
		]
		]
	\end{forest}
	\vspace{0.8cm}
	\caption{Folder structure of the dune-FRGDG module after initial setup. \vspace{0.4cm}\hspace*{\fill}}
	\label{fig:folderStructure}
\end{figure}
In the following we give step-by-step instructions to install the \emph{dune-FRGDG} module:\\
Before setting up the framework, all dependences of dune-FRGDG need to be available. These dependences are:
\begin{itemize}
	\item CMake $\geq$ 3.13
	\item Open MPI
	\item GCC with support for C++17
	\item pkg-config
	\item GSL - GNU Scientific Library
\end{itemize}
The framework itself can be cloned from the public github repository \cite{LDGGit},
\begin{lstlisting}[frame=single,basicstyle=\small,language=bash]
$ git clone \
    https://github.com/satfra/dune-FRGDG.git
\end{lstlisting}
Next, the user should run an automatic setup script and to build the framework itself:
\begin{lstlisting}[frame=single,basicstyle=\small,language=bash]
$ cd dune-frgdg
$ bash ./setup.sh
$ bash ./build-release.sh
\end{lstlisting}
The setup script will clone and modify all necessary DUNE modules for the framework. The script  \texttt{build-release.sh} will run CMake and make on all modules to produce a release build of the code.
Alternatively, if a debug build of dune-FRGDG is needed, the user can run \texttt{build-debug.sh} instead.
The header files, where most of the code is contained in, can be found at \texttt{dune-FRGDG/dune/FRGDG}. All source files are at \texttt{dune-FRGDG/src}.
After building the project, the binary executable files can be found in \texttt{dune-FRGDG/build-cmake/src}.
These can be either called single-threaded like
\begin{lstlisting}[frame=single,basicstyle=\small,language=bash]
$ ./anharmonicOscillator
\end{lstlisting}
or using MPI with
\begin{lstlisting}[frame=single,basicstyle=\small,language=bash]
$ mpirun -n 4 ./anharmonicOscillator
\end{lstlisting}
where one can replace the number 4 used here by the amount of desired MPI nodes.
\begin{figure*}
	\begin{subfigure}{0.65\textwidth}
		\centering
		\vspace{0.6cm}
		\begin{forest}
			for tree={
				align=center,
				font=\sffamily,
				edge+={thick, -{Stealth[]}},
				l sep'+=10pt,
				fork sep'=10pt,
			},
			forked edges,
			if level=0{
				inner xsep=0pt,
				tikz={\draw [thick] (.children first) -- (.children last);}
			}{},
			[Simulation
			[Simulation Set
			[Numerical Fluxes]
			[Model]
			[Grid]
			[FE Function Space]
			]
			[Driver
			[Scheme
			[Instationary PDE]
			[Stationary PDE]
			]
			]
			]
		\end{forest}
		\vspace{0.8cm}
		\caption{Logical structure of a simulation using the DUNE-FRGDG framework. The Simulation Set contains user-specifics, while the Driver performs the execution of all backend code. \vspace{0.4cm}\hspace*{\fill}}
		\label{fig:backendStructure}
	\end{subfigure}
	\hfill
	\begin{subfigure}{0.3\textwidth}
		\centering
		\begin{forest}
			for tree={
				align=center,
				font=\sffamily,
				edge+={thick, -{Stealth[]}},
				l sep'+=10pt,
				fork sep'=10pt,
			},
			forked edges,
			if level=0{
				inner xsep=0pt,
				tikz={\draw [thick] (.children first) -- (.children last);}
			}{},
			[main.cc
			[FRGDG.hh
			[simulationmgr.hh
			[simulation.hh
			[driver.hh]
			]
			]
			]
			]
		\end{forest}
		\caption{The execution flow of a typical program using the DUNE-FRGDG framework, starting from a file main.cc which contains the main function.\hspace*{\fill}}
		\label{fig:executionFlow}
	\end{subfigure}
\end{figure*}
%

\subsection{Implementation in DUNE}\label{app:outline}

Here, the basic design of the computational framework is outlined. The general structure is built on the basis of the dune-project \cite{dune24:16}, more precisely the \emph{dune-pdelab} module \cite{dune:pdelab}.
The framework is devised such that an end-user is only required to specify equations and some properties of the DG setup in a template. These specifications are then pieced together and used by the backend of the code to perform the simulation.

The general structure of the framework itself is shown in \Cref{fig:backendStructure}, the folder structure of the installed module can be seen in \Cref{fig:folderStructure}.
Any simulation consists of the user-specified simulation set (frontend) where all specifics of the simulation are abstracted, and the driver class (backend) where the actual simulation is performed.
The simulation set contains information about the chosen grid, finite element space, solver algorithms and the actual DG equations in the form of the numerical fluxes and a model class. The simulation set is further specified in \Cref{app:numset}.
The driver class uses a specified scheme that defines the way the equations are handled (in our case, the DG scheme and the LDG scheme) to do time-stepping. Within the schemes the way time-stepping is done is defined, i.e. for the DG scheme one uses a RK scheme to directly solve one instationary PDE, while the LDG scheme incorporates the solving of additional stationary equations into the RK scheme.

In practice, as any C++ program, the simulation starts with a main-function. Here, we first define a simulation which uses a specific simulation set (here for a large N-model) and then start the backend code execution flow by calling \texttt{Dune::startFRGSimulation}.
\begin{lstlisting}[frame=single,basicstyle=\small,language=C++]
using SIM = Simulation<largeN::SimSet>;
Dune::startFRGSimulation<SIM>(argc, argv,
	"largeN.ini");
\end{lstlisting}
The function \texttt{Dune::startFRGSimulation} is found in the header file \texttt{FRGDG.hh} and does some MPI setup, as well as reading the supplied initialisation file before calling the simulation manager \texttt{SimulationMgr}:
\begin{lstlisting}[frame=single,basicstyle=\small,language=C++]
SimulationMgr<SIM> mgr(ptree,helper);
mgr.run();
\end{lstlisting}
In the simulation manager a simulation is created and started from the supplied simulation template and initialisation parameters:
\begin{lstlisting}[frame=single,basicstyle=\small,language=C++]
SIM sim(ptree,
	mpihelper.getCommunication(),0,log);
sim.start();
sim.finish();
\end{lstlisting}

Finally, the simulation itself invokes the driver, which goes on to construct the DG system and solve it:
\begin{lstlisting}[frame=single,basicstyle=\small,language=C++]
DRV<N> driver(gv,fem,pP,ptree,mpicomm,log);
driver.Run();
\end{lstlisting}

The execution flow is shown in \Cref{fig:executionFlow}. Although the step through the \texttt{SimulationMgr} may seem superfluous, the structure is designed to allow extending the \texttt{SimulationMgr} to do massive parallelisation of simulations and their management.
%

\subsection{Numerical settings}\label{app:numset}

Before implementing a physical model we must specify the numerical setting, i.e. the grid, the integration scheme or the numerical fluxes. This is done in the \texttt{SimSet} class in the \textit{models/ON.hh} header file, which also contains the physical equations.

Generally, LDG methods are defined on a domain of arbitrary dimension composed of $K$ cells, $\Omega = \cup_{k=0}^K D_k$.
Therefore, we must chose an appropriate grid-implementation:
\begin{widetext}
\begin{lstlisting}[frame=single]
using RF = double;
static constexpr unsigned dim = 1;
using GridConstructor = CubicYaspGridConstructor<RF, dim>;
\end{lstlisting}
Here, we first set all computations to be performed with double precision and then specify a one-dimensional grid to be used. In principle the grid dimension can be set to any integer number. Currently, our computations use cubic grids, i.e. rectangular grid cells.
\\
The next step is to specify the test-function space within each grid cell $D_k$. A numerical approximation using test-functions $\psi_n$, compare \labelcref{eq:Approx}, to the exact solution ${\bf w}$ of \labelcref{eq:DGForm} reduces the degrees of freedom of ${\bf w}_h^k$ to a finite number. This test function space is set up in the following way:
%
\begin{lstlisting}[frame=single]
template<int order>
using FEM = Dune::PDELab::QkDGLocalFiniteElementMap<RF, double, order, dim,
	Dune::PDELab::QkDGBasisPolynomial::legendre>;
static constexpr std::array<int,4> orders {1,2,3,4};
\end{lstlisting}
%
The module \textit{dune-pdelab} \cite{dune:pdelab} provides the polynomial space $Q_k$, which is spanned by all polynomials up to order $k$ in the according number of dimensions, for example $Q_1 = \{1, x, y , xy\}$ if \texttt{dim=2}. The Legendre polynomials are chosen as a basis of this test-function space.
Also, when increasing the dimensionality of the grid, an appropriate tensor-product of one-dimensional function spaces is used. Furthermore the polynomial orders $\{1, 2, 3, 4\}$ are precompiled into the program and are available to chose from in the \textit{.ini} file. We note here that the code can be used for finite volume computations if the polynomial order is set to 0. Similarly, we recover a finite element method (FEM) if continuity is enforced between cells.\\
The numerical approximation from \labelcref{eq:Approx} is inserted in \labelcref{eq:DGForm} and projected onto the test function space. This formulation does not account for possible discontinuities at the cell borders. Worse, the solution is effectively double-valued at the cell borders $x_r^{k-1}=x_l^k$. This necessitates the introduction of a numerical flux:
%
\begin{lstlisting}[frame=single]
using Numflux = utils::static_switch<idx,LLFfluxLDG<Model<idx>>,
	LLFfluxLDG<Model<idx>>>;
\end{lstlisting}
\end{widetext}
This example uses our standard choice of numerical flux, which is the LLF-flux. Other possible choices are implemented and can be found in the \texttt{numericalflux} folder, such as up- and down-winding fluxes, as well as a central flux.

Now, the inside of each cell reduces to a finite element problem where the boundary conditions are given by the numerical flux.  An appropriate solver is the key ingredient to FEMs and is provided by the \textit{dune-istl} module \cite{dune-istl}. Here we distinguish the solvers given by the \texttt{LS\_stat} and the \texttt{LS\_instat} types, for the stationary and instationary equations respectively, in the LDG setting. The correct interplay of both solvers is ensured by the \texttt{LDGScheme}.
%

\subsection{The initialisation file}\label{app:inidat}

In this Section we comment on the parameters  which can be specified after compiling the program, in the initialisation file. The initialisation file is a source file and can be found in the \texttt{ini} folder. Running a build script will automatically copy the initialisation files to the location of the corresponding executable. The file is divided into several sections, firstly we start with the grid. Here, it remains to specify the length of the grid, as well as the cell density:
\begin{lstlisting}[frame=single]
 [grid]
 origin = 0
 L = 0.03
 N = 30
 localRefinement = 0
\end{lstlisting}
All grid parameters need to be specified for each grid dimension, i.e. the input for a two dimensional grid should be given as \texttt{N = 30 30}. \texttt{origin} specifies the lower left corner of the grid. \texttt{L} and \texttt{N} denote the grid length and the number of cells respectively.\\
The \texttt{localRefinement} parameter allows for a finer cell spacing in certain areas of the grid by bisecting all cells $n=\text{\texttt{localRefinement}}$ times.
This feature can be used by setting \texttt{localRefinement} to $n>0$ and specifying the additional parameters \texttt{localRefinement $n$}, e.g.:\\
\begin{lstlisting}[frame=single]
 [grid]
 localRefinement = 2
 localRefinement0 = 15
 localRefinement1 = 10
\end{lstlisting}
In this example, all evenly spaced cells of the original grid up to cell $15$ are bisected once and afterwards the cells up to (the original) cell $10$ are bisected once again.

Next, the finite element properties, i.e. the inside of single cells, are specified:
\begin{lstlisting}[frame=single]
 [fem]
 degree = 2
 torder = 3
\end{lstlisting}
\texttt{degree} sets the polynomial order in each cell. Here, the only available options are those previously specified in the simulation set. The order for the Runge-Kutta time-stepping method is set by the \texttt{torder} parameter. The currently available options range from first to fourth order RK methods.\\
All input parameters for the physical model are specified within the \texttt{param} section:
\begin{lstlisting}[frame=single]
 [param]
 Lambda = 0.65
 l = 1
\end{lstlisting}
Here, all initial UV-parameters, such as the initial cutoff-scale $k_{\mathrm{UV}} =\ $\texttt{Lambda} or classical values of the couplings, can be added.\\
Another important element of the computation is the RG-time integration:
\begin{lstlisting}[frame=single]
 [time]
 maxTime = 3
 maxTimeStep = 3e-3
 minTimeStep = 1e-15
 safetyfactor = 0.001
 timeGrid = 1e-2
\end{lstlisting}
Starting from $t_\mathrm{UV}=0$ one would ideally like to integrate up to $t_{\mathrm{fin}} \to \log(\frac{0}{\Lambda}) = \inf$. Since this is impractical we usually set a \texttt{maxTime} up to which the equations are solved. The final RG-scale is then given by $k_\mathrm{IR} =\Lambda \exp(- \texttt{maxTime})$.
We limit the maximal step size by \texttt{maxTimeStep} as a safety precaution. Similarly, we specify a minimal time step (\texttt{minTimeStep}), which causes the computation to abort if the step-size goes below this value. Usually, uncharacteristically small time-steps are caused by non-converging solutions. Therefore, we abort the program to avoid spending a lot of computation time on these computations. Ideally, the code limits the time step using the Courant-Friedrichs-Lewy (CFL) conditions \cite{CFL1967}. In case of diffusive systems the condition is usually enforced with an additional safety-factor of about \texttt{safetyfactor}$\approx 0.001$. Furthermore, we specify the rate at which the computed data should be saved by \texttt{timeGrid}.
Lastly, it is specified how the data should be saved:
\begin{lstlisting}[frame=single]
 [output]
 name = lN
 subsampling = 0
\end{lstlisting}
\texttt{name} is prepended to the computation-specific details, which are added to the name of the output-file in the model class. The \texttt{subsampling} option allows to increase the amount of data points saved. When setting this to zero, the saved data-points in the output files are minimally sampled according to the polynomial order. For smoother output this option can be increased.

%
\begin{figure*}
	\begin{minipage}[b]{0.4925\linewidth}
		\includegraphics[width=\linewidth]{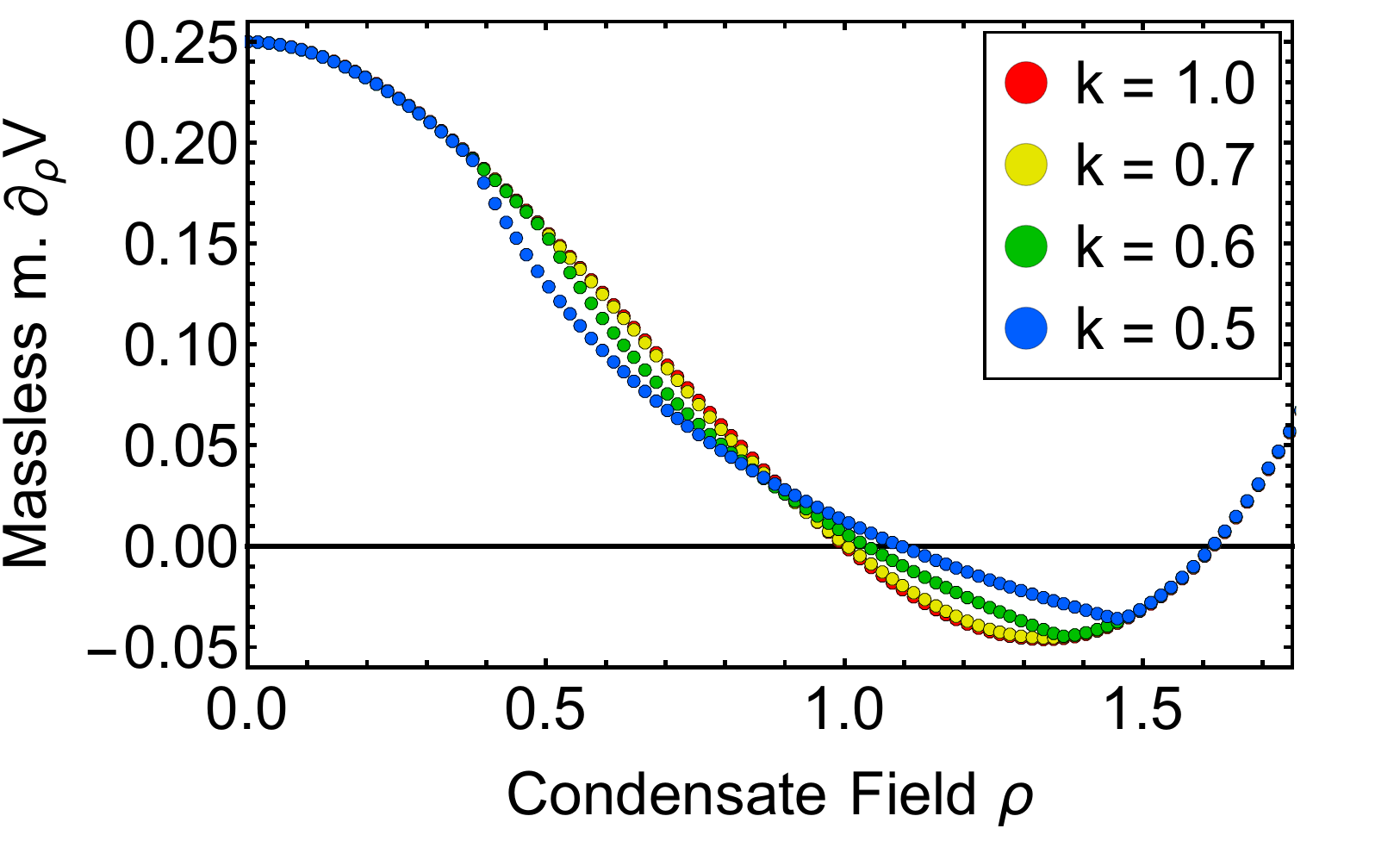}
		\subcaption{RG-time dependence of the massless modes $\pi$.\hspace*{\fill}}
		\label{fig:shock1}
	\end{minipage}%
	\hspace{0.01\linewidth}%
	\begin{minipage}[b]{0.4925\linewidth}
		\includegraphics[width=\linewidth]{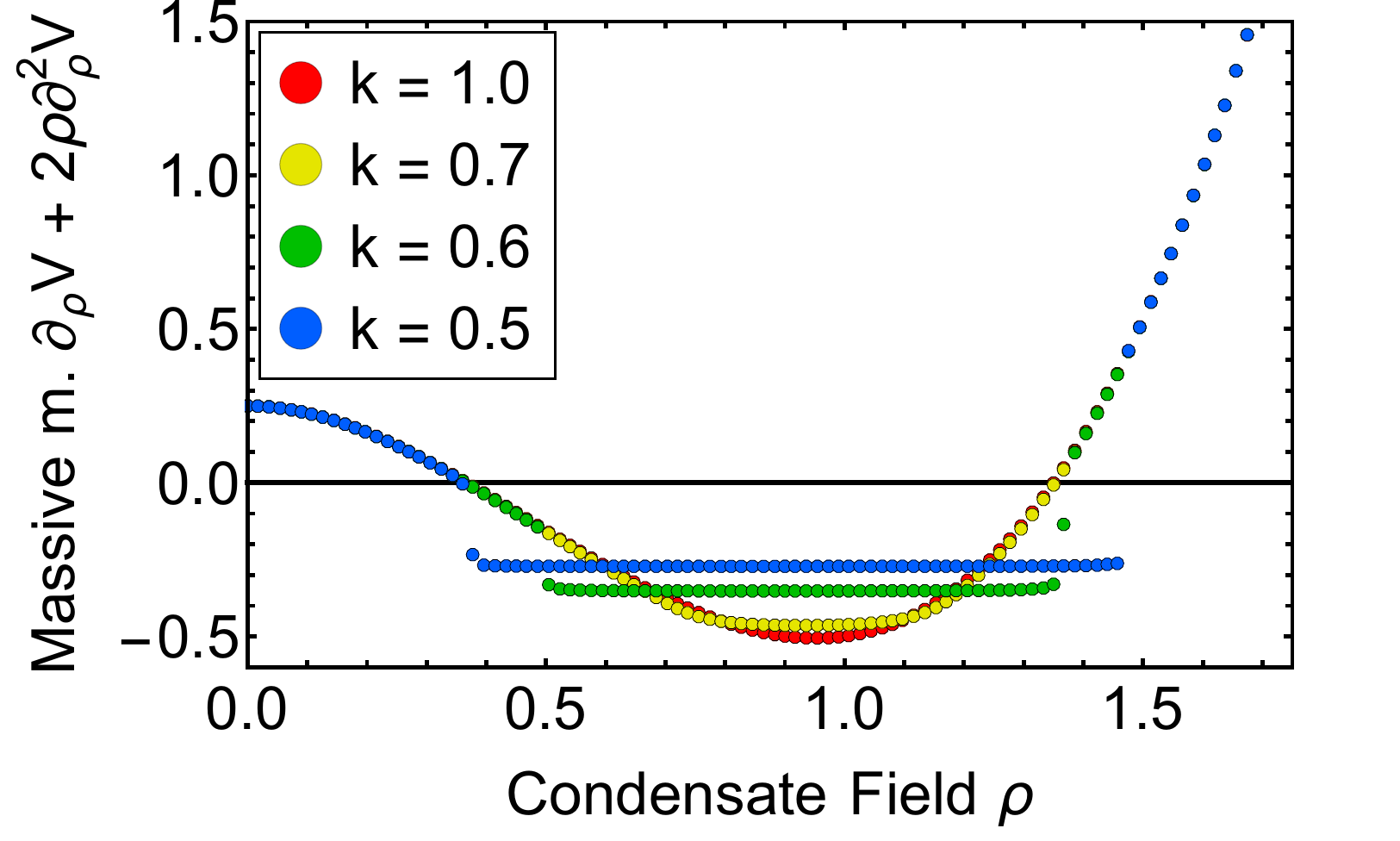}
		\subcaption{RG-time dependence of the massive mode $\sigma$.\hspace*{\fill}}
		\label{fig:shock2}
	\end{minipage}
	\caption{Shock development in the O(N) model at finite $N$. Shocks can appear in the massive mode $\sigma$ and translate to kinks in the massless mode $\pi$. The initial conditions for this computation are given in \labelcref{eq:iniShock} with an initial cutoff $\Lambda=1$. The equation is terminated at $k=0.5$ due to a very small step-size. \hspace*{\fill}}
	\label{fig:shock}
\end{figure*}
%

\section{Linearising the diffusion}\label{app:DGFlow}

The flow equation of the $O(N)$ model \labelcref{eq:DGflowqm} is a highly non-linear convection diffusion equation. Therefore, an application of the LDG method presented in \Cref{sec:NumericsLDG} requires a few manipulations of the equation. Linearity in the second derivative is achieved by computing the RG flow of the first potential derivative $\partial_\rho V_k(\rho)$ and separately of the second $\partial^2_\rho V_k(\rho)$, introducing artificially a second instationary equation. An according manipulation of the flow is detailed in the following.

Starting from the flow of the O(N) model \labelcref{eq:FlowEqFiniteN}, we obtain the flow of the pion mass $u = \partial_\rho V_k(\rho)=m_\pi^2$ by taking a $\rho$ derivative
\begin{subequations}\label{eq:flowDetails}
\begin{align}\label{eq:flowDetailsa}
	\partial_t u =& \partial_\rho ( f_{N-1}(u) + f_1(u+2 \rho v)) \,,
\end{align}
where the bosonic flux is given by $f_i(x)=A_d \frac{ i }{1+ \frac{x}{k^2}}$ and the prefactor $A_d=\frac{v_d k^{d}}{2(2\pi)^d}\,$.
The diffusive contribution is captured in $v=\partial^2_\rho V_k(\rho)$, which appears in \labelcref{eq:flowDetailsa} in a non-linear way. We can solve this issue by deriving a flow equation for $v$ explicitly. Taking an additional $\rho$ derivative of \labelcref{eq:flowDetailsa} yields
\begin{align}
	\partial_t v =& \partial_\rho \partial_t u  \notag\\[1ex]
	=& \partial_\rho \big(  \partial_\rho(f_{N-1}(u) +f_1(u+2 \rho v)) \notag\\[1ex]
	=& \partial_\rho \big(  v \ \partial_u f_{N-1}(u) +
	\partial_\rho s  \partial_s f_1(s)\vert_{s=u + 2 \rho v} \big) \,.
\end{align}
\end{subequations}
The last line now contains a simple convective flux and a second term which suits the LDG scheme presented in \Cref{sec:NumericsLDG}. Furthermore, the correct diffusive behaviour, i.e. a flattening and not a steepening of slopes, is ensured by the requirement $0>2 \rho \partial_s f_1(s)$. For a more throughout discussion see \cite{Cockburn98thelocal} and in a fRG context \cite{Ihssen:2022xjv}. In our case, the fulfilment of this requirement is always given since
\begin{align}
	\partial_s f_1(s) = - \frac{A_d k^2}{(k^2+s)^2} \,.
\end{align}
We continue by computing the flux of the stationary equation in \labelcref{eq:source}, where the integral can be computed analytically 
\begin{align}\label{eq:explicitstatflux}
	j(u+2\rho v)=& \int^{u+2\rho v}_0 ds  \frac{A_d^{1/2}k}{(k^2+s)} \\[1ex]
		=&   A_d^{1/2}k \log(k^2+(u+2\rho v)) - \mathrm{const}\,.
\end{align}
The constant drops out, since $j$ is a flux. Hence only the derivative, or differences of $j$ are evaluated in the computation.
The argument of the logarithm in \labelcref{eq:explicitstatflux} is always positive, due to the convexity restoring properties of the equation and therefore $k^2+(u+2\rho v)>0$ at all times during the flow \cite{Litim:2006nn}.

\section{Shock development}\label{app:shock}

We investigate the possibility of shock development in the O(N) model at finite $N$, in particular the physical scenario $N=4$. 
Shock development can be found in the large-$N$ limit of the O(N) model for specific sets of initial conditions \cite{Grossi:2019urj}. Furthermore the high density region of the quark-meson model in the large-$N$ limit also holds the potential for shock development \cite{Grossi:2021ksl}. Shock development in the quark-meson model is linked to competing order effects at high densities. Its occurrence is decided by the dynamics and respective strengths of the quark and bosonic fluxes. In the O(N) model it is much easier to tune the initial conditions such that shocks develop. The general idea is to create a non trivial minimum in the non-convex initial potential $V_{t=0}(\rho)$, which is given by
\begin{subequations}\label{eq:iniShock}
\begin{align}\label{eq:iniShock1}
	V_{t=0}(\rho) = \lambda_2 \rho + \lambda_4 \frac{\rho^2}{2} + \lambda_6 \frac{\rho^3}{3} + \lambda_8 \frac{\rho^4}{4} \,.
\end{align}
We now discuss the requirements of shock formation for different couplings $\lambda_i$.
The convective dynamics of the bosonic/scalar field are such that waves always propagate to the IR, i.e. to lower field values $\rho$. In the absence of diffusion, i.e the large-$N$ limit, a shock builds up in the massless mode/pion at some position $\rho_0 \neq 0$ if the flow towards the IR is sufficiently 'jammed' at some value $\rho<\rho_0$. This translates into the requirement $F(\rho) \ll F(\rho_0)$ for some $\rho<\rho_0$, where $F$ refers to the pion flux defined in \labelcref{eq:ONequations}. In the large-$N$ limit of the O(N) model, it is possible to create shocks in a $\phi^6$ potential, i.e only using the couplings $\lambda_2$, $\lambda_4$ and $\lambda_6$, while $\lambda_8=0$ in \labelcref{eq:iniShock1}, see \cite{Grossi:2019urj}.
At finite $N$, shock development becomes harder to grasp. The dynamics shift towards the massive mode/sigma. The sigma mass $m^2_\sigma = \partial_\rho V + 2 \rho \partial_\rho^2 V$ contains a second derivative term \labelcref{eq:masses}, which acts as a non-linear diffusion term in the pion-flow $F$. With this additional diffusive contribution the $\phi^6$ initial conditions are no longer able to produce shocks in the current setting. This scenario changes, if we include higher orders. Using the initial conditions
\begin{align}\label{eq:iniLambdas}
	\lambda_2 = 0.25\,,  \ \lambda_4 = 0\,, \ \lambda_6 = -0.5 \ \mathrm{and} \ \lambda_8 = 0.25\,,  \quad 
\end{align}
\end{subequations}
at $\Lambda=1$ we discover the formation of shocks in the massive mode, compare \Cref{fig:shock}. Interestingly, the non-linear diffusion term in the pion-flow $F$ allows for the formation of a kink, but not of a shock in the pion mass $m_\pi^2 = \partial_\rho V$. \Cref{fig:shock2} demonstrates nicely, how a kink in the massless mode $\pi$ translates to a shock in the massive mode $\sigma$. Due to the high numerical cost in the precise evaluation of shock development in convexity restoring flows, we terminate the computation in \Cref{fig:shock} at $k=0.5$ with a RG-time step-size $\le 10^{-10}$. Further investigations in this direction are deferred to further work and require technical improvements, such as the correct limiting procedures or improved time stepping schemes.

In summary, we can report on the appearance of shocks and non-analyticities in the O(N) model. The development of a shock is tightly linked to the form of the potential and requires the formation of a non-trivial minimum at finite RG-scale $k$, as well as higher order scattering processes at the UV cutoff. These conditions cannot be excluded in more elaborate models, for example low energy effective models of QCD at high densities. Hence, we hope to present an in depth analysis of shock development at finite $N$ in connection to first order phase transitions in future work. A similar studies have been done in \cite{Grossi:2019urj, Grossi:2021ksl} for the large-$N$ limit.
 
 \vfill
\begingroup
\allowdisplaybreaks


\endgroup


\bibliography{paper_refs}
\end{document}